\documentclass{article}

\usepackage{amsmath}
\usepackage{amsfonts}
\usepackage{amssymb}
\usepackage{graphicx}
\usepackage{subcaption}
\usepackage{wasysym}
\usepackage{url}
\usepackage{palatino}
\usepackage[biblabel]{cite}
\usepackage[hidelinks]{hyperref}
\usepackage{mdwlist}

\graphicspath{ {figures/} }

\newcommand{\optimist}{\noindent \ $\raisebox{-2pt}{\Large{\sun}}$}
\newcommand{\lunatic}{\noindent \ $\raisebox{-1pt}{\Large{\leftmoon}}$}
\newcommand{\etal}{\emph{et al.}}

\title{The Bayesian optimist's guide to adaptive immune receptor repertoire analysis}
\author{Branden J. Olson and Frederick A. Matsen IV}

\usepackage{fancyhdr}
\begin{document}
\maketitle

%
%

\begin{center}
Please cite as:

Olson, B and Matsen IV, FA. The Bayesian optimist's guide to adaptive immune receptor repertoire analysis. \emph{In press, Immunological Reviews.}
\end{center}

\vspace{1cm}

\begin{abstract}
Probabilistic modeling is fundamental to the statistical analysis of complex data.
In addition to forming a coherent description of the data-generating process, probabilistic models enable parameter inference about given data sets.
This procedure is well-developed in the Bayesian perspective, in which one infers probability distributions describing to what extent various possible parameters agree with the data.
In this paper we motivate and review probabilistic modeling for adaptive immune receptor repertoire data then describe progress and prospects for future work, from germline haplotyping to adaptive immune system deployment across tissues.
The relevant quantities in immune sequence analysis include not only continuous parameters such as gene use frequency, but also discrete objects such as B cell clusters and lineages.
Throughout this review, we unravel the many opportunities for probabilistic modeling in adaptive immune receptor analysis, including settings for which the Bayesian approach holds substantial promise (especially if one is optimistic about new computational methods).
From our perspective the greatest prospects for progress in probabilistic modeling for repertoires concern ancestral sequence estimation for B cell receptor lineages, including uncertainty from germline genotype, rearrangement, and lineage development.
\end{abstract}

%


\section*{Introduction}


\subsection*{How to read this paper}

\begin{itemize}
\item If you are an immunologist and want to learn more about probabilistic modeling, start here.
\item If you love probabilistic modeling and are curious about immune repertoires, you may want to start by getting background in immunology in general\cite{Sompayrac2011-fk} and immune repertoires in particular\cite{Robins2013-ww,Yaari2015-pc}, then reading the Models section.
\item If you already know both topics and get bored easily, skip to your favorite parts of repertoire analysis.
\end{itemize}

\subsection*{Why bother with probabilistic models?}
Before entering on our quest for model-based analysis of repertoires, one might ask ``why bother?''

The first answer is simple: repertoires are generated by a probabilistic process of random recombination, unknown pathogen exposures, and stochastic clonal expansion.
Thus, when analyzing repertoires it behooves us to reason under uncertainty.
The last century of statistical development offers a refined set of tools to make statements about such systems and assess our confidence in them.

Second, repertoire data shows us that complex models are justified.
For example, not all germline genes are used with equal frequency in repertoire generation.
The frequency of these germline genes is interesting to measure, but also informative of which genes were used in specific recombination events that gave rise to observed sequences.
Furthermore, the various genes all have characteristic distributions of trimming lengths, shown to be consistent between individuals\cite{Murugan2012-ue,Elhanati2015-ld,Ralph2016-kr}; incorporating this further improves annotation and clustering inference.
Such observations can also suggest mechanistic hypotheses that can then be tested with experiments.

Third, the probabilistic approach offers a principled means of accounting for hidden latent variables that form an essential part of the model, but are not themselves of direct interest to the researcher.
For example, we may not care about the exact rearrangement event that led to a given B cell receptor, but this is still an important latent variable for clustering analysis: indeed, one should only cluster receptors that came from identical rearrangement events.
Thus one can sum over the possible rearrangement events that led to this clonal family, leading to a natural means of evaluating a clustering likelihood\cite{Ralph2016-yl} that averages out uncertainty in the rearrangement process.

\begin{figure}
    \centering
    \includegraphics[width=.65\textwidth]{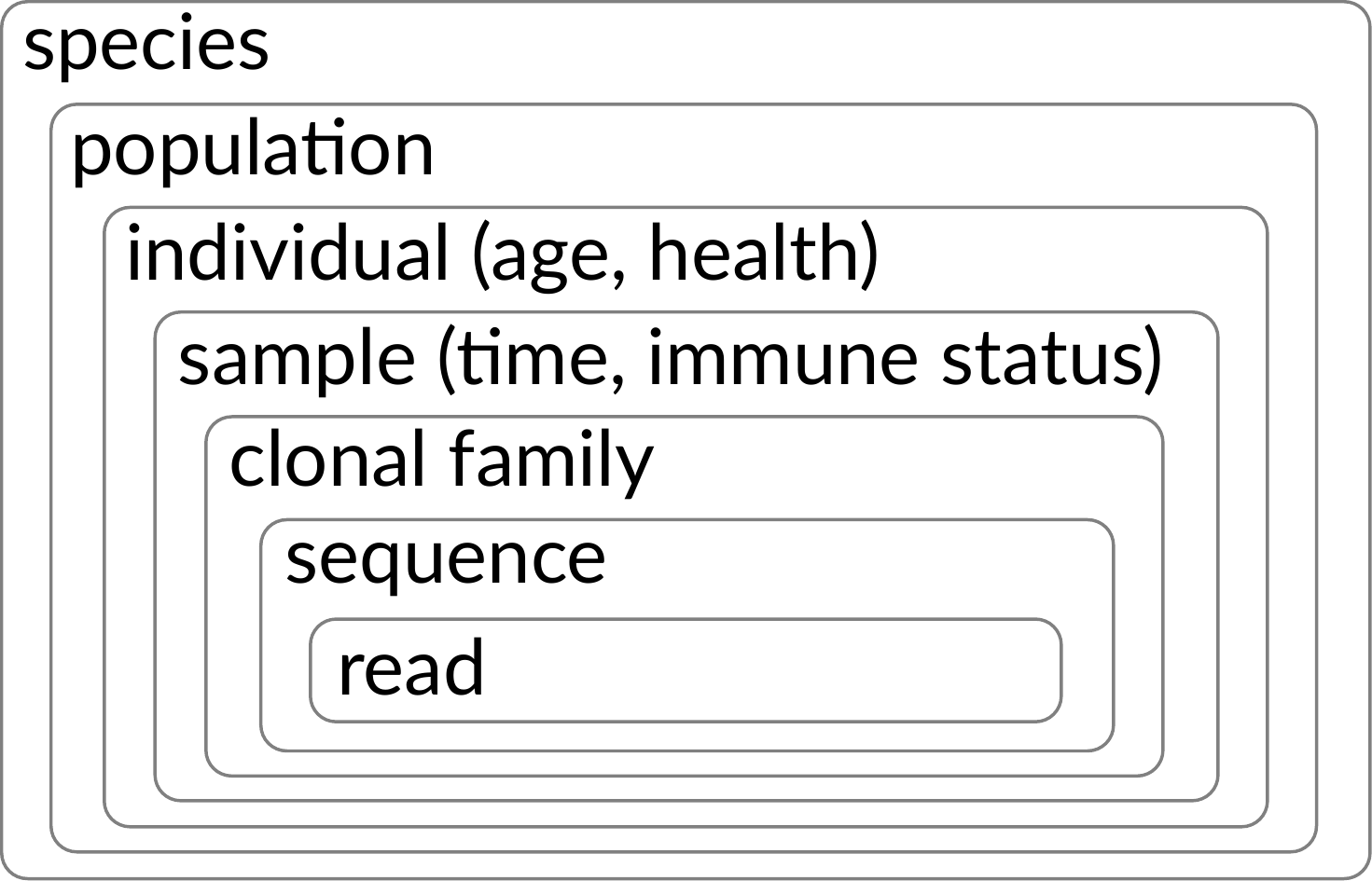}
    \caption{
    Immune repertoires are hierarchically structured, here illustrated by the hierarchy for B cell receptor sequences.
    We benefit by considering the whole hierarchy that contributes to our observable sequences rather than one sequence at a time.
    For example, by considering all the reads at once one can infer a personal germline set, which then informs the per-read annotation.
    By learning lots of personal germline sets one can infer population-level germline trends.
    }
\label{fig:hierarchy}
\end{figure}

Fourth, probabilistic models have well-developed notions of model hierarchy, in which inferences at each level inform and are informed by inferences at other levels.
This is essential to leverage the hierarchical structure present in immune receptor data (\emph{Fig.~\ref{fig:hierarchy}}).
For example, performing inference using many sequences at once (e.g. germline inference) can greatly improve per-sequence inferences, performing lots of per-individual germline inferences can tell us about the germline biases of a population, and so on up the hierarchy.

\subsection*{Model-based probabilistic analysis}
We begin by introducing model-based probabilistic analysis, and providing a very casual introduction to maximum-likelihood and Bayesian analysis as they apply to immune repertoires.

Consider a very simple model of the distribution of heights in a human population: a normal (a.k.a.\ Gaussian) distribution.
Say we have observed the height of all 127 million humans in Japan, rounded to the nearest centimeter, and we have plotted it as a histogram.
As a first approximation, one can think of fitting a probabilistic model as grabbing a normal distribution and flexing it with our hands until it looks as much as possible like that histogram.
If its estimates are too small, for example, we can scoot it right, and if it is too narrow we can bend it so it is broader.

\begin{figure}
    \centering
    \includegraphics[width=.95\textwidth]{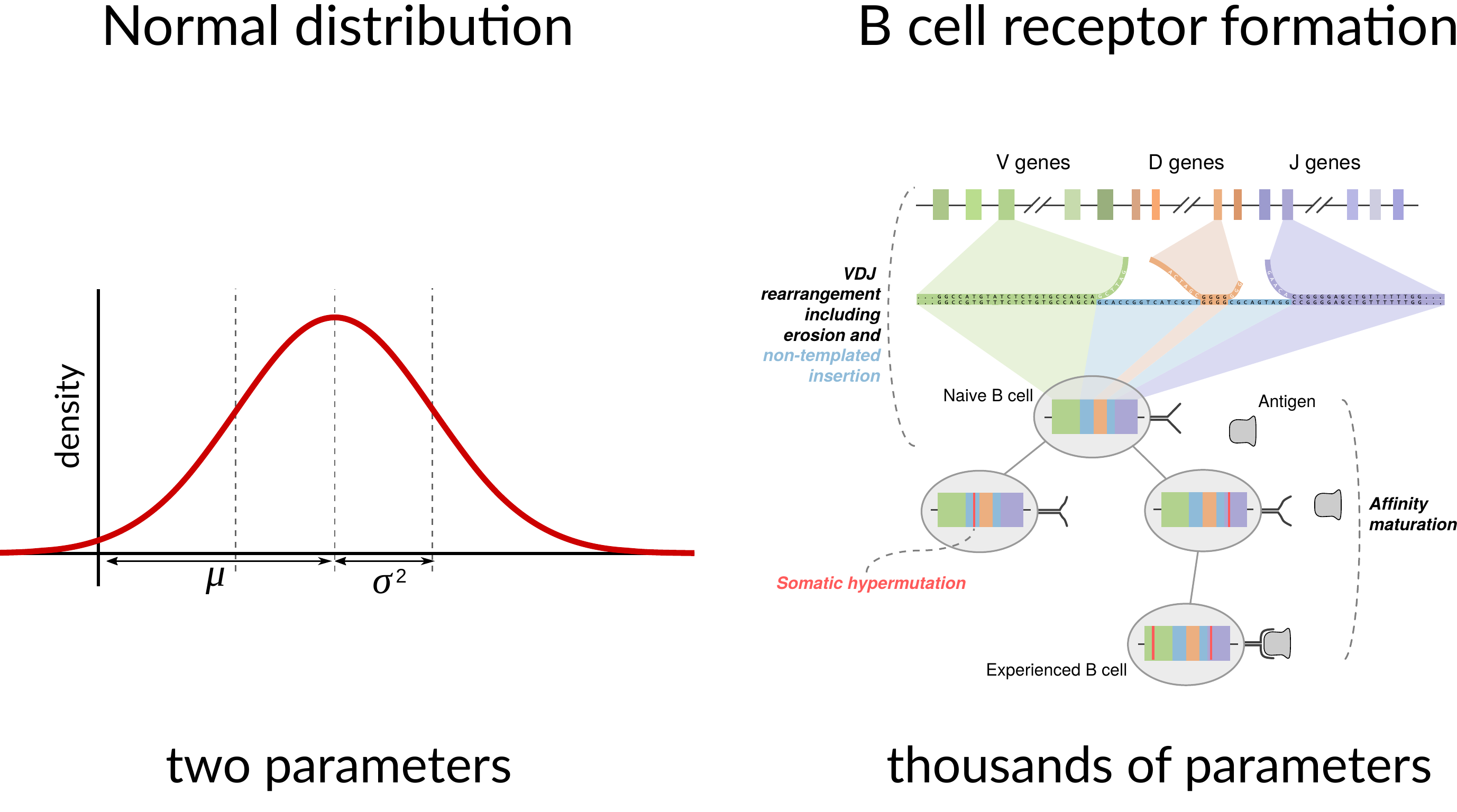}
    \caption{
    Human height and immune receptor formation can both be modeled using probabilistic methods.
    Right panel modified (with permission) from \cite{Murugan2012-ue}.
    }
\label{fig:probabilistic}
\end{figure}

This process can be formalized in terms of the principle of maximum likelihood, in which we find the parameter values that are most likely to have generated the observed data.
The likelihood function of the model parameters can again be thought of as ``the probability of obtaining the observed data under the given model with those parameters.''
Although not quite a rigorous definition for all settings, this definition is rigorous for discrete data such as heights rounded to the nearest centimeter, or DNA sequences.
For a normal distribution model, which is parameterized by mean $\mu$ and variance $\sigma^2$, we can directly calculate this likelihood function.
This likelihood is a product of terms, one for each human, equal to the Gaussian probability density $(2 \pi \sigma^2)^{-1/2} \exp[-(x-\mu )^{2} / 2\sigma ^{2}]$, where $x$ is the height of that human.
It turns out that the usual formulas for the mean (i.e. the sample average) and the (biased) sample variance (the average squared deviation from this mean) are exactly the maximum likelihood values of these parameters: those values that maximize the likelihood function!

One can also approximate the likelihood function using repeated simulation for a single set of parameters, as we illustrate using the following thought experiment.
In the heights example, we can estimate the likelihood as follows: generate many samples of size 127 million from a normal distribution rounded to the nearest integer, and calculate the fraction of times we get exactly the observed set of heights.
Although this will be an extraordinarily small number, it will be larger for parameter values that fit well (the normal distribution fits the histogram of measurements closely) than for ones where it does not, and thus is a means of doing parameter fitting.
[Note that these two perspectives on optimizing our model, that of picking model values such that simulation is as close as possible to observation, and that of maximizing a likelihood function, are actually identical if we define ``as close as possible'' in terms of Kullback-Liebler \cite{MacKay2003-dm} divergence.]

The inferential setup is the same for immune repertoire analysis, except that the models and data are more complex (\emph{Fig.~\ref{fig:probabilistic}}).
Rather than having a model that generates human heights, models for immune repertoire analysis generate immune repertoires as collections of DNA sequences.
In a similar way, fitting such models is a process of wiggling the parameters until the model generates repertoires that are as similar as possible to the observed repertoires.
In certain cases we can efficiently compute a likelihood function such that optimizing this function does the wiggling more formally, using either an exact formula or approximate numerical routines.
However, this is not the case for all models, and indeed much of the subject of the second half of this paper describes models that attempt to make a balance between computability and realism.

For repertoires, we can again imagine maximum likelihood fitting happening via simulation: we have a model from which we can simulate repertoire sequences, and we can approximate the likelihood for a collection of parameters for a given data set based on the fraction of times it generates the observed data exactly.
In principle, we can fit the model by iteratively wiggling parameters and re-simulating, preferring those wiggles that more frequently generate the same data as what was observed.
Of course, for real data sets such fitting is sheer lunacy: a repertoire simulation will never once exactly match an observed repertoire sample containing a million unique sequences even if we were to run it for our whole lifetimes!
Nevertheless, this is a helpful thought experiment that underlines the importance of likelihood functions, which can be thought of as a ``short cut'' avoiding such simulation.
We will continue this thought experiment below.

We can continue the height metaphor to explain Bayesian analysis.
Bayesian analysis again concerns model parameters $\theta$.
In the heights example, $\theta$ is a pair consisting of the mean $\mu$ and the variance $\sigma^2$.
The goal of Bayesian analysis is to not just find the best parameters $\theta$, but to get an ensemble of possible values of the parameters along with an idea of how well each describes the data $x$.
This is formalized in the notion of a posterior distribution, which is a probability distribution on the collection of parameters describing how likely the various parameters are to be correct given the data.
Having a full distribution over parameters rather than just point estimates allows for a more detailed characterization of the uncertainty in our inferences.
For example, we can summarize this posterior distribution in terms of credible intervals, which are the Bayesian analog of confidence intervals.
To obtain such a posterior distribution for our height example, we begin by specifying a prior distribution on the parameters.
This prior is our \emph{a priori} idea of what the heights might be before we sample any data.
We then incorporate the data to get a posterior distribution.
Formally, this comes from the deceptively simple statement of Bayes' theorem
\[
p(\theta \mid x) \propto p(x \mid \theta) \, p(\theta)
\]
which states that the posterior distribution $p(\theta \mid x)$ of model parameters $\theta$ given data $x$ is proportional to the likelihood $p(x \mid \theta)$ times the prior $p(\theta)$.
We can think of this as re-weighting our prior assumptions based on how well they explain the data.

This sounds simple enough, but in fact thousands of careers of computational Bayesians have been dedicated to the challenge posed by Bayes' theorem being expressed in terms of proportionality rather than equality.
Indeed, even if we can say how much one parameter set is better than another via Bayes' theorem, we have to evaluate many different parameters to obtain a value for the posterior, which makes a statement about how good a given parameter set is compared to \underline{all} possible parameters.
The situation is analogous to that of climbers in a mountain range tasked with estimating their height relative to the average height of the range: it is easy to see that one location is higher than the other, but evaluating the average height requires traversing the entire range and taking careful measurements.
This is an informal way of saying that the integral of the posterior distribution is typically intractable.

When the posterior integral is in fact tractable, as can be the case for very simple models,  we can obtain the posterior distribution directly as a formula.
In our height example, if we take a normal prior distribution for the mean with a fixed variance, we can directly obtain a formula for the posterior distribution (which turns out to also be normal).
However, such directly-computable models with so-called ``conjugate priors'' are few and far-between, and none of them involve immune receptor DNA sequences.

For more complex models we do not attempt to compute the posterior distribution directly, but rather we sample from it.
In this way we obtain a ``histogram'' that approximates the full posterior distribution: in our heights example, we would get a collection of $(\mu, \sigma^2)$ samples from the joint distribution on these parameters.
It is common to summarize these samples in terms of their single-variable posterior estimates, which in our example would be one histogram for the mean of the height distribution and another for the variance.

Although sampling from posterior distributions is a challenging problem, decades of research has developed sophisticated methods, as well as probabilistic programming languages that are dedicated to the task \cite{Hornik2003-fu,Lunn2009-sv,Carpenter2016-dn}.
We will briefly summarize one method (MCMC) below, but first present a completely rigorous but utterly impractical means of sampling from a posterior distribution via simulation.
In the above thought experiment, we were approximating the value of the likelihood for a single set of parameters, and here we have an even more ambitious goal: to approximate the posterior across parameter values.
Repeat the following process to obtain a posterior sample on parameters given some data:
\begin{itemize*}
\item draw values of the parameters from the prior
\item simulate data using those parameters
\item does this simulated data match the observed data exactly?
\item if so, add these parameters to our posterior sample, and if not discard them
\item return to the first step until the desired number of samples is obtained
\end{itemize*}
In the height example, each such cycle involves drawing 127 million samples from a normal distribution and checking if they are the same as the observed data.
The result is a sample from the posterior distribution on $\mu$ and $\sigma^2$.
In an immune repertoire example, we could do the same by simulating sequences, which is even less practical than the completely impractical idea of applying this to the height example.
Luckily, there are other means of sampling posterior distributions.

The most common method for sampling from a posterior distribution is Markov chain Monte Carlo (MCMC).
MCMC is a random procedure that moves around parameter space such that the frequency with which the procedure visits a given parameter is proportional to its posterior probability.
The most popular type of such inference in phylogenetics is random-walk MCMC\cite{Nascimento2017-wf}, in which parameter values (such as a tree topology and its branch lengths) are perturbed randomly; these perturbed values are always accepted if they are ``better'' and accepted with some probability if they are ``worse.''
Being able to accept ``worse'' parameter modifications is important so that the algorithm explores the entire space rather than getting stuck at the peak of a distribution.
The notions of ``better'' and ``worse'' are determined by the Metropolis-Hastings ratio, which depends on having a likelihood function that can be evaluated efficiently.
This sort of sampling is implemented in packages such as BEAST \cite{Drummond2012-mh} and MrBayes \cite{Ronquist2012-hi}, but due to computational complexity is typically limited to hundreds of sequences in a single tree.

Before exploring computational challenges, we describe marginalization and discuss priors.
Marginalization is the practice of ``integrating out'' nuisance parameters, which are parameters that are important for the model but may not be of interest for the researcher.
Imagine we were interested in what D gene was used for a given B cell receptor sequence, and want to take a probabilistic approach because such assignment is naturally uncertain.
In a likelihood-based approach, one can only evaluate the suitability of a D gene assignment when we also have specified the amount of trimming encountered by this D gene, even if that parameter is not actually of interest to us.
Therefore we sum over the possible amounts of D gene trimming.
In general this is called integration because summation is a special case of integration.

Prior distributions require careful consideration.
All distributions, including prior distributions, have parameters that must be chosen.
The parameters of prior distributions are called ``hyperparameters.''
Where do those come from?
One option is to use a hierarchical Bayesian analysis in which we consider prior parameters as random variables themselves, also requiring prior distributions.
The phylodynamics community have developed sophisticated methods to infer mechanisms of viral spread using such a hierarchical approach\cite{Dudas2017-sb}.
However, at some point this recursion must end and one must either fix values arbitrarily or attempt to estimate them from the data.
The process of estimating fixed hyperparameters is known as empirical Bayes \cite{Efron2012-go}.

\subsection*{An informally-described hierarchy of inferential difficulty}

Here we describe a difficulty hierarchy for maximum likelihood and Bayesian inference based on how difficult the model is to compute.
\begin{enumerate}
\item \textbf{Conjugate priors available for model:} In this case, the posterior is available as an exact formula.
Hence, no sampling is required, and the posterior is extremely efficient to evaluate.
Unfortunately, this is never the case for repertoires.

\item \textbf{Efficiently computable likelihood function available:} Here, maximum likelihood estimation is tractable, and Bayesian methods can be used via Markov chain Monte Carlo (MCMC).
Phylogenetic trees under models where each site evolves independently fall into this category, as the Felsenstein algorithm\cite{Felsenstein1981-zs} provides a means for efficient likelihood evaluation.
Nevertheless, tree inference is still challenging, and provably hard (in the technical sense) given difficult data \cite{Roch2006-ub} because of the super-exponential number of trees that must be tried in order to be sure of finding the best one.

Repertoire analysis methods such as hidden Markov models (HMMs, described in more detail below) for rearrangement inference also fall into this category.
In this case there is also latent state (that is, the transition points between germline sequences and the N/P junction between germline-encoded regions); this latent state can be efficiently marginalized by the Forward-Backward algorithm\cite{Durbin1998-uq,Murugan2012-ue}.
Bayesian estimation for such parameters is also possible \cite{Scott2011-zf} though has not been applied to repertoires.

\item \textbf{Efficiently computed likelihood function available if we condition on some additional latent state}
Some models do not have an efficiently-computable likelihood function in general, though a likelihood can be computed if we expand the parameters of interest to include some additional information.
For example, the ideal phylogenetic reconstruction method for repertoire data would take the context-sensitive nature of somatic hypermutation into account \cite{Rogozin1992-xv}.
We can efficiently compute a likelihood function using a context-sensitive model such as S5F \cite{Yaari2013-dg} if we specify the order of and time between mutations. However, these additional parameters are not typically of interest and thus need to be marginalized out using Markov chain methods \cite{Wei1990-yq}.
For certain classes of such models, only the order of mutations (versus their exact timing) matters \cite{Feng2017-dh}.

\item \textbf{No likelihood function available}
When no likelihood function is available one must resort to simulation-based methods such as approximate Bayesian computation (ABC) \cite{Marin2012-gt}.
In this method, one obtains approximate posterior distributions by reducing the data to relevant summaries and seeing which models produce data that match these summaries well.
Our above thought experiment required an exact match of simulated and experimentally-derived data in order for a set of parameters to be accepted.
In ABC, one accepts parameters with a probability determined by how closely pre-specified summary statistics of the simulated data agree with those of the experimental datasets.
This has been applied with success in population genetics problems with a modest number of parameters.
However, as the model complexity grows, even simulation-based methods suffer the ``curse of dimensionality'' and will eventually become intractable.

Any sufficiently detailed model of repertoire generation will land here.
For example, it's not possible to calculate likelihoods for complex models based on agent-based simulation \cite{Robert2017-ki}, although one could sample them using ABC.
In fact, an informal version of ABC is currently used in  B cell receptor sequence analysis, in which one adjusts simulation parameters until they generate data that looks close to experimental data according to a battery of summary statistics\cite{Kleinstein2003-mu,Magori-Cohen2006-wd,Shahaf2008-cc}.
\end{enumerate}

We see that there is often a balance between realism and computability; although there is no inherent reason why this must be so, it is often the case.
For example, computation is eased by assuming variables in a model are independent, even if that's not exactly true.
In the above hierarchy, this is illustrated by easy-to-compute site-independent phylogenetic models on one hand versus hard-to-compute context-dependent models on the other.

\section*{Models}

Here we describe existing and potential probabilistic models for immune receptor development.
Although in principle any probabilistic model can be used for inference (via the ``thought experiment'' inference procedure described above), we find it useful to distinguish between inferential models and models for simulation.
For the purposes of this paper, inferential models are those that are meant to be fit to data to learn something about the underlying system.
We will be interested in inferential models that are tractable to use for inference if one is ``optimistic'' (marked with \optimist): at least, one should be able to do inference on each individual component using existing machinery.

Models for simulation serve a separate and essential purpose.
Such models can be more complex and need not have an efficiently-computable likelihood to be useful.
Agent-based models, such as models of a germinal center\cite{Robert2017-ki} fall into this category.
Models can make predictions, such as the groundbreaking 1993 prediction of cyclic re-entry\cite{Kepler1993-xj} that was dramatically validated over a decade later\cite{Victora2010-zg,Mesin2016-ew}.
Also, if we want to validate inferential algorithms, we need accurate generative models.
For these reasons we are going to sketch ``lunatic'' model components (marked with \lunatic) as well, for which we only require the ability to simulate in forward time.

We will investigate this framework while following receptor development from the germline gene repertoire to clonal expansion.
For every component of the process, we will follow an identical pattern in this order: biological background, then previous work on inference, then sections on ``optimist'' \optimist\ and ``lunatic'' \lunatic\ models.
The biological background will of course be a miniscule fraction of what is known, as we can only include parts that are relevant for the modeling goals here.

\begin{figure}
    \centering
    \includegraphics[width=.65\textwidth]{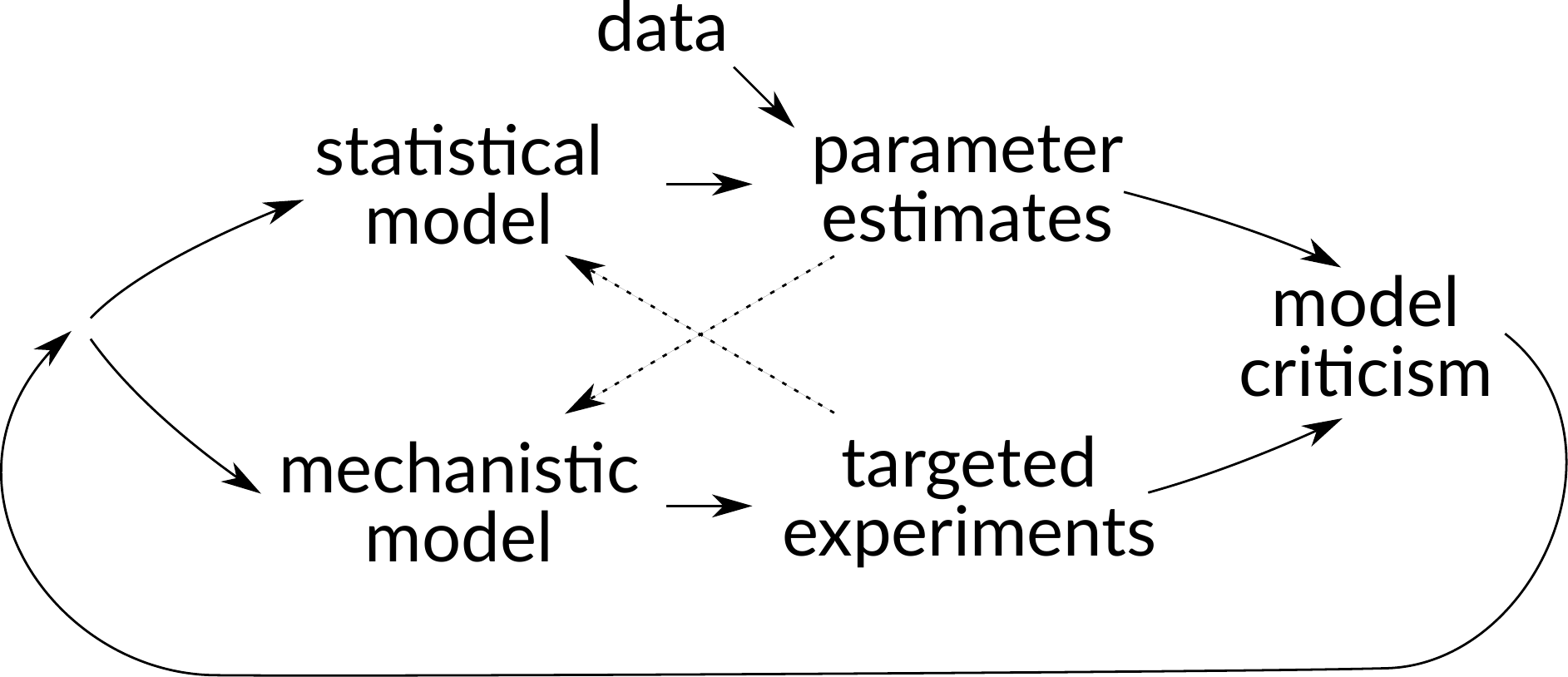}
    \caption{
    Statisticians and biologists have typically approached adaptive immune receptor research in two parallel tracks.
    Statisticians (upper path) treat data as given and perform model criticism based on numerical estimates of how well models fit the data.
    Biologists (lower path) formulate their models mechanistically and generate data in targeted experiments to directly test their model.
    An integrated approach (dashed arrows) has biologists using parameter estimates to formulate mechanistic models, and statisticians using the results of targeted experiments to formulate statistical models.
    Ideally, the distinction between these two classes of models would evaporate, although mechanistic models are not always readily fit using statistical means.
    }
\label{fig:box}
\end{figure}

Before we begin this voyage, we note that traditionally biologists and statisticians have slightly different but not incompatible notions of what is meant by ``model.''
A biologist's model is typically a conceptual model describing the mechanistic process by which something happens.
For example, transcription factor $X$ binds cofactor $Y$ which allows it to initiate transcription of gene $Z$.
Such a model may not have any parameters and thus cannot ``generate'' data, although it can typically be used to devise an experiment to test the hypotheses of the model.

A statistician's model, on the other hand, need not have any mechanistic underpinning, although it necessarily contains parameters and can be used to generate data.

These two perspectives lead to different means of iterative model improvement (\emph{Fig.~\ref{fig:box}}).
Biologists scrutinize their models for components that can be separated out and perturbed individually to form a test of the model.
This reductionist approach has taught us most of what we know about biology today.
Statisticians, on the other hand, are generally interested in evaluating models via model fit.
That is, if we generate data from our model, does it resemble our observed data, and are new, unseen data values described well by the model?
If not, how can we add model components that will result in a better-fitting model?
This iterative process of model improvement has been called ``Box's loop.''\cite{Blei2014-wn}

Nonetheless, these viewpoints are quite compatible, and indeed we may need to combine them to meet the next set of challenges in adaptive immune receptor research.
For the statistician, incorporating mechanism into statistical models means that inferred parameters have direct interpretation, and such models typically have fewer parameters.
For the biologist, formalizing a biological model statistically means that models with hidden parameters can be directly compared in a rigorous way.

With all this introduction out of the way, now we begin considering probabilistic models for adaptive immune repertoires!

\subsection*{Germline genotype}
Although repertoire modeling often starts with V(D)J rearrangement\cite{Bassing2002-bf}, the rearrangement process is in turn determined by the genotype of each individual, in particular the collection of germline V, D, and J genes in the loci forming the various receptors.
A complete survey\cite{Watson2012-ea,Jackson2013-ad} is out of the scope of the paper, but suffice it to say that although complete haplotypes at germline loci have been sequenced \cite{Rowen1996-pb,Boysen1997-bp,Watson2013-rm,Watson2014-op}, the genetic diversity of these loci is high and many new alleles continue to be discovered, especially in non-Caucasian populations \cite{Feeney1996-ea,Wang2011-mk,Gadala-Maria2015-co,Scheepers2015-gs}.
Thus, the germline genotype forms an important part of the hidden state for repertoire generation\cite{Watson2017-mr} with real medical consequences\cite{Feeney1996-ea,Avnir2016-vs}.

This motivates inference of germline genotypes directly from repertoire sequence data.
Early work performed such inference by carefully considering several sets of high-throughput sequencing data~\cite{Boyd2010-yj,Kidd2012-bt,Elhanati2015-ld}.
Kidd \etal\ used a maximum likelihood model assuming uniform gene use to infer alleles, and were able to phase these onto haplotypes for individuals who are heterozygous at \emph{IGHJ6}\cite{Kidd2012-bt}.
Later work used naive sequences and an assumption of no gene duplication to iteratively obtain haplotypes via probabilistic gene assignment for three individuals\cite{Elhanati2015-ld}.
More recent work has delivered automated tools for germline set inference: TIgGER\cite{Gadala-Maria2015-co}, IgDiscover\cite{Corcoran2016-yy}, and partis\cite{Ralph2017-ih}.
TIgGER introduced a ``mutation accumulation'' plot relating mutations at a given site to the overall level of mutation in sequences.
This plot should have a smooth shape in the absence of new alleles, but a ``bend'' in the presence of an unannotated allele; partis works by explicitly searching for this bend.
IgDiscover applies hierarchical clustering to naive-sorted data to obtain germline sets in species for which little or no germline information is known.

\optimist\
There is a substantial need for probabilistic germline repertoire inference methods.
For example, this would be very useful if we want to estimate the unmutated ancestor of B cell clonal families while quantifying uncertainty.

Germline gene inference is inferred from a whole repertoire at a time; this fact alone poses some challenges to a probabilistic method.
For example, if we wish to compare two germline gene sets, naively one would need to perform a complete re-alignment of all sequences to obtain a likelihood.
Because such a likelihood is available, it can be formally classified as ``efficient'' (second category in the above hierarchy) although such repeated re-alignment is not practical for large data sets.
Some cleverness would be helpful here, such as only re-annotating sequences that could be affected by changing the germline set.
Another alternative would be to infer a too-large pool of possible candidate germline genes, perform probabilistic alignment using all of these germline sequences, and use the associated probabilities without running complete re-alignment in a second step to cut down the pool.

Haplotype inference has been shown to be a useful tool for cutting down germline sets from such a candidate pool\cite{Elhanati2015-ld,Kirik2017-qq}.
This works by assuming limited or no gene duplication of individual genes in a germline set, and then using joint gene usage under VDJ recombination to infer which alleles lie on which haplotype.
One can then review the gene assignments and reject suspicious inferences, since having many alleles of a given gene inferred to lie on a single haplotype casts doubt on their authenticity.
This method does have some caveats.
It requires heterozygosity at J (or D) genes, and that the V gene in question is expressed at reasonable levels on both chromosomes.
Also, the immunoglobulin germline locus is dynamic with many gene duplication and conversion events, so we cannot exclude the possibility of many alleles of a gene being present on a haplotype.

Despite these caveats, in order to have the best germline gene inference it may be useful to extend inference to the full pair of haplotypes, called a ``diplotype.''
In order to do so we will need to formalize generative models on diplotypes, which could act as a prior for inference.
Ideally, such a generative model would come from observing many diplotypes.
As noted above, such direct haplotype sequencing is rare, but this situation may change using improved assembly techniques applied to long-read sequencing data.
Alternatively, one could build up such a model by taking a large ensemble of data sets and iteratively estimating haplotypes and prior parameters (determining, e.g.\ the prior distribution of the number of alleles per gene) using empirical Bayes.
A parameterized prior could be developed based on racial background, in which people with genetic ancestry from various places would have a different distribution of germline genes.
The biological importance of gaining broad diplotype information has been carefully laid out in a recent review\cite{Watson2017-mr}.

In principle one could directly use probabilistic methods to infer a pool of possible germline sequences, although here the problem of requiring re-annotation becomes much more acute because of the many hypotheses that must be tried.
The TIgGER mutation accumulation plot gives the values of a complex conditional probability, and at least for the near term it makes sense to continue using this summary.
Its analysis could be improved by more flexible models of how mutations accumulate on sequences--- current efforts implicitly assume that each site accumulates mutations linearly as a function of the total number of mutations on that sequence.

\lunatic\
Germline genotypes differ between individuals in a population and between populations and species because of long-time-scale evolutionary processes; it is already tempting to work to better understand these processes of mutation and selection.
In terms of mutation, one may wish to connect germline gene change with mechanistic models of gene duplication and loss\cite{Reams2015-bo}.
Analysis of large populations\cite{Luo2017-gp}, especially using parent-offspring data\cite{Luo2016-kg}, will be important to develop such models.
It is also tempting to infer selection on germline gene sets to maintain a diverse pool of starting material for VDJ rearrangement, as well as to make it easier to mount an antibody-mediated response against locally-important pathogens.
Much more data, in particular data spread across many more populations, will be required to perform such inference.

\subsection*{Rearrangement}

By ``rearrangement'' we mean joint gene choice, trimming, and insertions in the process of V(D)J recombination\cite{Schatz2011-cm} without any selective steps for tolerance or binding.
Biologically, this is determined at least in part by gene location, presence of recombination signal sequences \cite{Cowell2003-vh}, chromatin accessibility of these sequences\cite{Schatz2011-cm}, and long-range loop structure \cite{Montefiori2016-pw}.
There are surely additional complex genetic determinants of the rearrangement process, and how those contribute to repertoire formation will be a continued topic of research.

This complex machinery leads to a complex probabilistic process that determines rearrangement \cite{Jackson2013-ad}.
Prior work has found interaction between N nucleotide addition and recombination\cite{Kepler1996-kd} as well as dependence between D and J gene use in BCRs\cite{Volpe2008-oq}.
There is also clear evidence of interaction between gene use and trimming length\cite{Murugan2012-ue,Elhanati2016-yq}.
The rearrangement processes change with age, a phenomenon recently quantified in mouse\cite{Sethna2017-yl}.
In addition to the usual rearrangement process, oddities such as VH replacement\cite{Wilson2000-bg,Collins2004-hw,Meng2014-zw} and inverted and multiple D genes do occur, although recent analyses indicate that these are rare in the overall repertoire\cite{Ohm-Laursen2006-yq,Lee2016-tt}.

This process is greatly deserving of complex models, as all variables determining the rearrangement process are both interesting and decidedly non-uniform.
Indeed, many probabilistic models have been formulated, with the hidden Markov model (HMM) framework being particularly popular\cite{Ohm-Laursen2006-yq,Gaeta2007-mz,Munshaw2010-mj,Elhanati2016-yq,Ralph2016-kr}.
Various implementations of the HMM differ in the parameterization of gene choice and trimming distributions, with the trend being towards parameter-rich categorical distributions for trimming.
Such rich distributions are justified by the observation that although trimming distributions are different between genes, strong concordance between individuals shows that the models are not simply fitting noise\cite{Elhanati2016-yq,Ralph2016-kr,Marcou2018-sm}.
Recent work has extended this to a more general modeling framework expressible in terms of an arbitrary Bayesian network \cite{Marcou2018-sm}.
For the insertion sequences, applying an HMM has shown dependence of the next base on the previous one\cite{Elhanati2016-yq}.

\optimist\
Despite substantial progress, there is still work to be done describing the rearrangement process using probabilistic models.
The distribution of inserted sequences invites further exploration: are more complex models warranted?
Although current models are inferred per-data-set, it would be helpful to have models that can concern multiple related data sets and parameterize differences between them using covariates such as age \cite{Sethna2017-yl}.
Such models may also be useful to infer differences in the rearrangement process by genetic background \cite{Feeney1996-ea,Kidd2015-vt,Avnir2016-vs}.
An obvious if formidable next step is to extend current probabilistic models to include complex rearrangements such as replacement and multiple D genes.

\lunatic\
It will be more challenging to relate these descriptive statistical models to mechanism.
As described above, gene choice is determined by recombination signal sequence strength and accessibility.
Sequence features must also govern the amount of trimming, and early work found sequence motifs that change the distribution of trimming amounts\cite{Nadel1995-rv,Nadel1997-lv}.
This work has not been extended in our current era of abundant high-throughput sequencing data sets.
However, our biological knowledge has also expanded: we now know gene choice is determined by processes including megabase-scale loops and chromatin state, although the roles of various processes such as hairpin opening and nucleotide deletion to germline gene trimming are still something of a mystery.
Since these processes are so complex, any proposed model will have to judiciously choose a balance between realism and tractability.
Also, such a project will require diverse expertise in statistical modeling and biological mechanism.

\subsection*{Initial selective filters}
Positive and negative selection determines which B and T cells are able to circulate.
Positive selection ensures that T cells are able to bind major histocompatibility complex (MHC) molecules of the host.
Negative selection happens to avoid self-reactivity and maintain an appropriate level of interaction with MHC.
In B cells the initial selective processes ensure that a functional antibody is produced with limited self-reactivity.

For B cells, previous work has found selection against long and/or hydrophobic HCDR3 loops \cite{Larimore2012-lo} as well as selection on germline gene use \cite{Meng2011-ht} and D gene frame\cite{Benichou2013-ej}.
Others have inferred ``selection factors'' for various aspects of the TCR\cite{Elhanati2014-mf} and BCR\cite{Elhanati2015-ld} in the initial selective process.
These factors are multiplicative terms that describe the probability of seeing a sequence with a specific characteristic in the post- vs pre-selection repertoire.
For example, a selection factor for a specific amino acid at a specific location quantifies the level to which this amino acid is selected for or against.

\optimist\
There are still many possible ways to extend analysis of ``bulk'' characteristics of sequence-level selection.
Current methods analyze the role of a single feature at a time, such as CDR3 length or particular selection factors, and it would be interesting to look for selection on sets of factors.
Paired heavy/light and alpha/beta chain data is also an interesting source for such joint selection analysis\cite{DeKosky2013-iz,Howie2015-vp,Grigaityte2017-dp}.
It will also be important to take MHC type into account when performing such analysis and look for MHC-mediated effects, although the largest analysis so far found relatively few TCRs that were negatively associated with MHC \cite{Emerson2017-co}.

\lunatic\
It will be very difficult to make the leap from such a ``bulk'' analysis of sequence-level selection to the true prize, which is inference on a per-sequence level.
Extracting the binding properties of a BCR or TCR from sequences, and hence its potential for autoreactivity, is a grand challenge of computational biology that will not be solved soon.
Perhaps the best strategy will be via protein structural modeling, or via machine learning techniques applied to large data sets of pre- and post-selection receptor sequences.
Thus such per-sequence analysis appears to be out of scope of the sort of probabilistic modeling considered here.

\subsection*{T cell clonal expansion}
T cells are stimulated to divide when they bind to an MHC loaded with a peptide that they recognize.
This is called clonal expansion.

As in the previous section, one can again consider two questions for clonal expansion: first, what are bulk characteristics of the expanded repertoire, and second, can we infer anything on individual TCR sequences?
Regarding bulk characteristics, abundance distributions of T cells have proven to be a fertile means of learning about the patterns of antigenic stimulus and competition\cite{Desponds2016-ja,Desponds2017-kj}.
Twin studies show a strong genetic effect of T cell clonal expansion in terms of overall memory cell response and response against a specific immunization\cite{Qi2016-or}.
T cell development changes through age, which has been used to show that our naive T cell repertoire is a complex mixture of cells generated at different ages\cite{Pogorelyy2017-gx,Sethna2017-yl}.
Aging clearly modifies both the existing repertoire and our capacity to respond to novel stimulus\cite{Boyd2013-lt}.

Regarding individual sequences, it is interesting but difficult to associate characteristics of specific TCR sequences with genetics and immune state.
One component of this is to develop relevant notions of similarity between receptors, which can then be used to perform clustering and projection into a lower dimensional space\cite{Dash2017-fr,Yokota2017-rr}.
Using these and related tools, recent work has moved towards a variety of machine learning goals, including
clustering sequences according to their specificity using tetramer binding data\cite{Dash2017-fr,Glanville2017-uw},
predicting new sequences that will bind a given epitope\cite{Glanville2017-uw},
identifying relationships between TCR sequence and MHC use\cite{Sharon2016-ib,Emerson2017-co},
and finding sequences or sequence characteristics that differ between groups\cite{Thomas2014-jo,Emerson2017-co,Ostmeyer2017-rz,Pogorelyy2018-ox}.
Building databases of epitope-TCR pairs\cite{Shugay2017-nj} and high throughput measurement of affinity\cite{Birnbaum2014-xw} will certainly spur this development.

\optimist\
Probabilistic modeling can be used to estimate the chance of obtaining a given TCR sequence, and thus has an important role in interpreting T cell frequency data.
Such models can be used to estimate the degree of antigen-stimulated clonal expansion by comparing the probability of generation to the TCR frequency \cite{Pogorelyy2018-ox}.
In addition, probabilistic modeling is appropriate for interpreting bulk properties of repertoires in terms of clonal relationships within the sequences.
Here, we will certainly see continued development of models describing the abundance distribution of T cell clones, and how these change with age and immune stimulus.

On the other hand, predicting biophysical properties of individual TCR sequences is not easily solved using a model-based probabilistic framework.
The binding fitness landscape of individual sequences is too ``rough'' for typical probabilistic methods--- a small modification in the right place can take a strongly binding receptor and make it non-binding.

\lunatic\
Thinking about a generative model that is too complex for inference, we seem to fall between two stools: probabilistic generation models and models of bulk properties seem tractable, whereas models of individual sequences seem impossible or inappropriate with current probabilistic tools.

\subsection*{B cell clonal expansion: antigenic stimulus}
B cell clonal expansion is complex and delicious for statistical modeling, and thus we will divide our description of this process among the following five sections.

In response to immune challenge, dense accretions of lymphocytes form structures known as germinal centers \cite{Victora2012-lu}.
These are the sites of B cell diversification, to which B cells gain entry by the ability to bind antigen.
This diversification process includes mutation and selection processes that will be covered in subsequent sections below.
After responding to the original infection, memory B cells and plasma cells are exported from the germinal center; after export these cells are mostly dormant until further stimulated.
Serum antibody levels are maintained by dynamically-controlled cell populations\cite{Manz2005-lq}.

Methods are now emerging to predict antigen-antibody affinity from sequence data for specific antigens\cite{Galson2015-ey}.
One approach is to find shared sequence characteristics.
For example, convergent sequence characteristics such as gene usage, CDR3 length, and mutation load have been identified in response to some vaccines\cite{Jackson2014-rx}.
However, not all sequence characteristics will have straightforward correlations: for example, an influenza vaccination experiment with closely sampled time points did not see a correlation between VJ gene usage and degree of expansion\cite{Laserson2014-yh}.
Age is an important covariate of this sort of analysis, as it changes the degree of hypermutation\cite{Jiang2013-kj}, how frequently certain gene combinations are generated in the unselected repertoire, and how gene combinations are selected in the memory repertoire\cite{Martin2015-dx}.

Another approach is to build out databases of BCR sequences responsive to specific antigens and look for similarity between them, in terms of both sequence similarity and time dynamics of re-activation\cite{Galson2015-cp,Galson2015-qk,Galson2016-iq}.
There is some, though not plentiful, data with which to infer these patterns, and the most information is available for HIV antigens\cite{Scheid2009-ot,Yoon2015-ec}.
This sort of approach may be aided by improved modeling of the space of antigen-binding sequences, such as with maximum-entropy models\cite{Asti2016-uy}, or some other means of predicting binding similarity using sequence information.

The effectiveness of a database-matching approach depends on similar BCR sequences being used to bind a given antigen, which leads to the subject of ``public'' repertoire analysis.
This sort of analysis determines which sequences are shared between individuals due to common antigen exposure and relatively high-probability random sequence generation~\cite{Galson2015-yl,Henry_Dunand2015-hf,Truck2015-aj}.
To make sense of the public repertoire one must understand the extent to which genetics and negative selection determine the naive repertoire.
For example, vaccination of human twins gives rather different results\cite{Wang2015-on}, while there is significant evidence of genetic predetermination for vaccination in mouse\cite{Greiff2017-hi}.
Indeed, it has recently been proposed that this represents a fundamental difference between the immune systems of these two species\cite{Collins2017-te}.

One can also consider various bulk properties of the memory BCR repertoire, and consider the difference between the naive repertoire and the mature repertoire.
A deep sequencing study observed differences in gene usage and CDR3 length\cite{DeWitt2016-yy}.
In addition, using appropriate lab and computational strategies, one can quantify and model the respective abundance distributions\cite{DeWitt2014-sz,DeWitt2016-yy}.

\optimist\
Like TCR sequences, probabilistic models are needed in public repertoire analysis of BCR sequences to disentangle the roles of similar rearrangements and antigenic stimulus in generating similar or identical receptors.
Analogously, models of abundance distribution should inform us about the dynamics of generation and selection, although there doesn't appear to be much work in this area yet.
Representations of sequences such as maximum-entropy models may provide useful tools for characterizing groups of antibody sequences binding a given antigen.

Although methods with more of a machine learning flavor may be a better fit for inferences on individual sequences, it may be useful to combine probabilistic models of sequence generation with probabilistic models of sequence families binding a certain antigen.

\lunatic\
As for T cell receptors, we again fall between two stools.

\subsection*{B cell clonal expansion: somatic hypermutation}
When B cells replicate, their BCR locus is mutated at a rate about a million times higher than in normal replicating cells.
This process is orchestrated by a complex set of steps, starting with deamination of a cytosine to make a uracil, and then proceeding down one of multiple paths of error-prone repair\cite{Methot2017-gi}.
These steps lead to complex context dependence, determining which antibodies are reachable via somatic hypermutation\cite{Hwang2017-tt}.

Several decades of work has focused on how these context ``motifs'' change mutability, first by finding ``hotspot'' motifs that are especially mutable \cite{Rogozin1992-xv,Dunn-Walters1998-ds,Cowell2000-rq} and then later by developing more quantitative approaches to describe the influence of various sequence characteristics.
This includes models estimating the mutability of all possible sub-sequences of some length \cite{Pham2003-jm,Yaari2013-dg,Cui2016-wz} or models that use sequence position and/or presence of individual bases at specific distances to predict mutability\cite{Cohen2011-rs, Elhanati2015-ld}.
Our group has recently generalized these approaches into a penalized survival analysis framework that can combine arbitrary sequence features, omitting those which do not  clearly contribute to improved model fit\cite{Feng2017-dh}.
An alternative way to formulate somatic hypermutation is to consider substitution frequencies of the combined mutation and selection processes on germline genes\cite{Ralph2016-kr,Sheng2017-ib,Kirik2017-bc,Dhar2018-pi}, although this does not provide predictions for N-region nucleotides.

Accurate mutation rate estimation is important for interpretation and prediction of B cell evolutionary patterns.
This is clearly true for estimation of natural selection\cite{Dunn-Walters1998-ra,Yaari2012-kk,McCoy2015-qi}, in which the mutation rate (the rate of introduction of nucleotide changes) is compared to the substitution rate (the rate of such changes that persist in the population) in order to estimate a natural selection parameter.
It is also important for understanding which antibodies are accessible from a certain rearrangement\cite{Hwang2017-tt}; analysis of B cell evolution over long timescales suggests decreasing mutability through time \cite{Vieira2018-yy}.

Currently there is an unfortunate division between biologically-based mechanistic models and statistical models, which are so far only descriptive.
Although the pattern of mutations has been used to state qualitatively that certain factors are important in the SHM process\cite{Dunn-Walters1998-ds,Spencer1999-tl,Rogozin2001-pw,Wilson2005-qa,Wang2010-lf}, this has not resulted in rate estimates for the various repair pathways.
The one exception is a mathematical model of AID activity in terms of scanning and catalysis\cite{Mak2013-iu}, although other processes are essential to the somatic hypermutation process \emph{in vivo}\cite{Chahwan2012-gp}.
A more mechanistically-explicit model of DNA damage and local error-prone repair should generate locally correlated sets of mutations more effectively, a task at which current models fail \cite{Marcou2018-sm}.

In addition to the point mutations described above, somatic hypermutation also introduces insertion-deletion mutations, or indels \cite{Wilson1998-br}.
The rate of indel introduction is comparable to the rate of point mutation\cite{Briney2012-sv,Bowers2014-no,Yeap2015-nl} although most of these indels are filtered out by natural selection in the functional repertoire.
Because the mechanisms for point and indel mutation are linked\cite{Methot2017-gi}, it is perhaps not surprising that correlation can be found between their locations\cite{Yeap2015-nl}.
Although most indels are filtered out, some have important functional consequences, such as in the development of broadly-neutralizing antibodies to HIV\cite{Kepler2014-pg}.

\optimist\
Inference of point mutation models is just starting to use methods with a probabilistic foundation, and more work needs to be done.
One outstanding challenge is that the process of somatic hypermutation happens on phylogenetic trees, and it is difficult to do model inference on phylogenetic trees with context-sensitive models.
Indeed, phylogenetic model inference typically integrates out potential internal states as part of the model fitting process on the tree; this is enabled by the use of the Felsenstein algorithm which requires an independence-among-sites assumption (more details below).
That assumption is of course violated for context-sensitive models.
Our group has used an additional sampling step to marginalize out the possible ancestral sequences of a given sequence, and avoid the need to do so in a fully phylogenetic context by selecting only one sequence per clonal family (i.e.\ phylogenetic tree).
Such estimation has been previously done for simpler classes of models\cite{Hwang2004-pj,Hobolth2008-dx}.

We are not aware of any probabilistic models for indels specifically in the somatic hypermutation process.
For molecular sequences in general, such models first appeared in 1986 \cite{Bishop1986-to}, followed by the foundational TKF models\cite{Thorne1991-fm,Thorne1992-nk}.
Recent work has defined a class of indel models with attractive computational properties \cite{Bouchard-Cote2013-ax,Zhai2017-ue}.

\lunatic\
A more biologically explicit mutation model would consider AID deamination and repair processes in terms of mismatch repair, base excision repair, indel introduction and gene conversion\cite{Methot2017-gi}.
Although a fully specified mechanistic model would be challenging for efficient inference, it is certainly suitable for simulation.
Our group is currently using our more flexible mutation modeling setup\cite{Feng2017-dh} to ``fish out'' certain types of effects that will allow us to estimate rates of these various pathways, and form the foundation for such models.

\subsection*{B cell clonal expansion: lineage development}
B cells undergo a Darwinian process of mutation and selection in the germinal center to improve binding to antigen.
Each B cell entering the germinal center founds a lineage (realized as a phylogenetic tree), and is the unmutated ancestor of all of its mutated descendants.
Mutation and selection happen in the dark and light zones of germinal centers, respectively: in the dark zone B cells reproduce, introducing additional diversity by the somatic hypermutation process described above, while in the light zone B cells compete to retrieve antigen from follicular dendritic cells.
Recent work has emphasized the importance of T cell help from retrieving antigen as opposed to direct stimulation to reproduce from BCR crosslinking\cite{Victora2010-zg}.

Affinity maturation is a dynamic population-level process.
Using a mouse engineered to express a reporter of apoptosis, researchers have found that apoptosis is the ``default'' outcome in the absence of T cell help\cite{Mayer2017-ij}.
Intensive examination of individual germinal centers has led to the hypothesis of ``clonal bursts'' in which B cells divide in rapid succession due to a strong T cell stimulus\cite{Tas2016-lq}.
Despite what would seem to be a very strong selective environment, phylogenetic analysis combined with affinity measurements has not revealed a steady march towards increased affinity in sampled germinal centers\cite{Kuraoka2016-zs,Tas2016-lq}.
Existing antibodies and B cells, including those appearing during the germinal center reaction\cite{Zhang2013-yn} and those from prior exposures\cite{De_Bourcy2017-ki}, change the evolutionary dynamics of the germinal center reaction.

Germinal centers are not seeded by single naive cells.
Indeed, random florescence labeling shows that many cells initially seed germinal centers, although these germinal centers often ``resolve'' to the descendants of a single cell through competition\cite{Tas2016-lq}.
In addition, B cells entering the germinal center need not be naive: mathematical simulation\cite{Or-Guil2007-vd} and mutation analysis of vaccination studies in mice\cite{McHeyzer-Williams2015-un} support the hypothesis that lineages can be re-seeded from existing lineages.

So much previous work has been done analyzing B cell sequence lineage development, that we will divide this section into further mini-sections.

\paragraph*{Clonal family inference}
Many computational methods have been developed to reconstruct the hidden aspects of B cell clonal expansion and infer the dynamics behind it.
Any bulk sample of B cells mixes sequences deriving from different naive cells and responding to different antigens.
Thus, an important first step for analysis is to group sequences into ``clonal families,'' namely, collections of sequences that descended from a single naive cell.
The most popular means of doing this is to apply single-linkage clustering to the sequences, allowing sequences to cluster if they are annotated to have the same V and J sequences, have the same CDR3 length, and are less than some fixed Hamming distance apart.
Needless to say, there are issues with each of these assumptions.
Somatic hypermutation may cause uncertainty as to germline gene assignment, and insertion/deletion mutations (indels) may change CDR3 length.
However, the assumption of a fixed cutoff, even a per-repertoire fixed cutoff\cite{Gupta2017-cg}, seems the most problematic.
The most obvious counter-example to this assumption is given by broadly-neutralizing antibodies against HIV, which with around 100 mutations have the same order of divergence from germline genes as these germline genes have to one another \cite{Wu2011-yy}.
From a phylogenetic perspective, fixed-cutoff methods make the surprising assumption that branch lengths in the process of somatic hypermutation cannot be longer than some fixed quantity.
This assumption seems even less sensible when we consider that repertoires are small samples from a large population; when we drop leaves from a phylogenetic tree because of sampling, the resulting branches become longer.

To avoid such assumptions, our group has developed a likelihood-based means of inferring clonal families in our \textsf{partis} software package\cite{Ralph2016-yl}.
We begin by recasting the problem to one of inferring groups of sequences that have the same naive sequence.
This differs from the original question of inferring clonal families, because the same naive sequence can be generated by two different rearrangement events.
To solve this question, ideally one would do a perfect job of inferring a naive sequence from each mature sequence and then simply cluster based on those inferred naive sequences.
However, such a procedure is not possible because there are many ways to obtain a given sequence from different ancestors via somatic hypermutation.
For this reason, the method calculates a likelihood that two groups of sequences come from the same naive ancestor, while integrating over possible naive sequences.
By comparing this likelihood to the alternative hypothesis that the two groups do not share ancestry via a likelihood ratio, one is able to decide whether these two groups should be merged into one.
The method applies this likelihood-based framework via agglomerative clustering in a manner reminiscent of the neighbor-joining algorithm\cite{Saitou1987-sr,Gascuel2006-gb}.
A naive implementation of this procedure would be far too slow for actual use, and thus, the method uses many optimizations.
Some of these come without any drop in accuracy, whereas some strike a balance between computational tractability and accuracy.

Inferences of such clusters will always be an uncertain process, which invites a Bayesian approach to obtain posterior distributions on the clusters.
Indeed, early unpublished versions of our procedure did this hierarchical agglomeration via sequential Monte Carlo (SMC)\cite{Doucet2001-av}, an algorithm that can be thought of as a probabilistically-correct type of genetic algorithm.
In SMC one maintains a population of objects being inferred, and at each stage makes some modification.
In this case, our software maintained a population of different partial clusterings, and at each stage every partial clustering makes some probabilistic merge weighted by a likelihood ratio.

This procedure was too slow and cumbersome to be applied to large sequence data sets.
However, our group experimented with it enough to feel confident that uncertainty was basically ``one-dimensional,'' such that the primary unknown quantity was the degree of clustering.
Given that partis records the sequence of clusterings that lead to each inferred cluster along with their likelihoods, we left our Bayesian ambition there.
A Bayesian clustering algorithm based on a Dirichlet process mixture model has been described\cite{Laserson2012-pi} although this algorithm does not appear to have been applied in practice.

\paragraph*{Phylogenetic inference on B cell sequences}
Even once the clusters are fixed, estimating the tree for each cluster is non-trivial.
Besides the fact that estimating a phylogenetic tree is an inherently hard problem, B cell sequences have features that differentiate them from typical applications of phylogenetics, and thus require special algorithms.
When sampling is dense, it is not unusual to sample ancestor-descendant pairs.
(Even if we aren't actually sampling the true ancestor of a given cell we may sequence a cell that is identical to it.)
The relatively short branch lengths between sequences has motivated an extensive use of parsimony\cite{Barak2008-fw,Stern2014-ph}, a method in which one chooses the tree that minimizes the number of mutations required to explain observed sequence data at the tips.
It is important to restrict the use of parsimony to cases with short branch lengths as it is known to be statistically inconsistent when branches become long\cite{Felsenstein1978-rr}; that is, it will produce the incorrect tree with probability one in the limit of long sequences.

When single-cell sequencing is applied to a densely-sequenced sample, such as one from a germinal center\cite{Tas2016-lq}, each sequence comes equipped with a meaningful abundance.
This information can be productively used to guide phylogenetic inference\cite{DeWitt2018-el}.
The intuition behind this approach is that, first, sampled abundance reflects the overall abundance of that genotype in the population, and second, more frequent cells are more likely to leave mutant descendants.
For this reason, we should prefer trees that connect descendants to more frequently observed ancestors over those that do not.

Substantial information about the ancestral sequence can be inferred with knowledge of germline sequences.
Indeed, if one knows that a given V gene was used in the process of VDJ recombination, we know the ancestral state for that region of the sequence.
This contrasts most applications of phylogenetics, in which ancestral states are typically unknown.
In order to integrate this information one needs a computational framework that knows about both VDJ rearrangement and phylogenetics.
Kepler has described such an approach, which iteratively infers a tree while estimating a posterior on unmutated ancestor sequences\cite{Kepler2013-sy}.
Each iteration takes the unmutated ancestral sequence with the highest posterior probability, builds a tree using that sequence at the root, and then re-estimates the posterior on the unmutated ancestor.

The highly context-sensitive mutation processes found in somatic hypermutation (reviewed above) violate the near-universal phylogenetic assumption of independent evolution between sites.
This assumption is essential for efficient likelihood computation in phylogenetics via the Felsenstein algorithm\cite{Felsenstein1981-zs}.
This can be understood intuitively as follows: if the substitution history at the first site depends on the second, and the history at the second depends on the third, then continuing this string of dependencies means that we must consider evolution to be happening on the whole sequence at a time.
This is computationally intractable as the state space for nucleotides is four to the power of the sequence length, defeating the traditional use of transition matrices.

Thus if one wants to stay inside the usual likelihood-based framework for phylogenetics one must use approximations to maintain the independence assumption.
An important step forward was recently made by incorporating context information into a codon model\cite{Hoehn2017-ol}.
In codon models, one considers codons, rather than individual nucleotides, to be the units of evolution and assumes independence between those codons.
By averaging out the part of nucleotide contexts that extend beyond the codon boundary, this work maintains a model that has independence between codons.
This approach has the additional advantage that one can estimate parameters of selection and context sensitivity directly from the model.

B cell sequence analysis has more emphasis on phylogenetic ancestral sequence inference than is typical for other applications of phylogenetics, and for good reason.
Ancestral sequence inference methods enable a beautiful convergence of computational analysis and laboratory experiments: estimated ancestral sequences can be expressed and built in the lab to test their properties\cite{Wu2011-yy,Zhu2013-ax,Doria-Rose2014-vi}.
Such experiments, when combined with structural analysis, give real insight into how substitutions lead to improved affinity.
The computational tool for these analyses has typically been PHYLIP\cite{Felsenstein1989-dt}, although other programs\cite{Ashkenazy2012-vk,Nguyen2015-bs} are faster or have additional features.

\paragraph*{Selection inference}
We can get additional insight into the evolutionary process by estimating the strength of natural selection on a collection of sequences using codon-based methods.
Such methods make inferences by considering the relative rate of synonymous (between codons for an amino acid) to nonsynonymous (between amino acid) substitution.
The intuition is that if there is selection to preserve an amino acid, one will see an excess of synonymous changes compared to nonsynonymous ones because nonsynonymous changes will be selected out of the population.
The opposite will hold for cases when amino acid change is beneficial.

Such analysis is made difficult by the context-sensitive mutation process: because the probability of substitution is influenced by the local sequence context on one hand, and natural selection on codons on the other, false conclusions can be drawn if one does not correct for it explicitly\cite{Dunn-Walters1998-ra}.
Such correction is indeed possible\cite{Hershberg2008-rp,Uduman2011-ib,Yaari2012-kk,Yaari2015-ss,McCoy2015-qi}.
Repertoire-level selection has been measured in the CDR region versus the framework region\cite{Hershberg2008-rp,Uduman2011-ib,Yaari2012-kk} and in the ``trunk'' (edges leading from the naive ancestor to the most recent common ancestor of sampled sequences) versus the rest of the tree\cite{Yaari2015-ss}, with results broadly consistent between individuals.
This theme of consistency is even more striking on a per-codon level\cite{McCoy2015-qi}, which shows diverse amounts of selection at various sites in the framework region that are consistent among individuals.

Tree shape and structure have also been used to estimate selective pressure on ensembles of trees.
Early work used graph-theoretic properties of trees to estimate selection strength\cite{Dunn-Walters2002-cu}; correlation between these measures and selection strength was determined by simulation \cite{Shahaf2008-cc}.
Later authors found that such properties can be distorted by difficult-to-control experimental factors \cite{Uduman2014-pb}.
They proposed an alternative method mapping mutations onto the edges of the tree and using patterns of replacement and silent nucleotide substitutions filtered to only include substitutions on non-terminal branches \cite{Uduman2014-pb}.

A more ambitious goal is to estimate selection on a single tree at a time.
One recent approach compares tree balance (the number of descendants on one side of a node versus another) at nodes directly below edges with amino acid changes versus those without\cite{Liberman2016-qh}.
An investigation of vaccine-responsive trees\cite{Horns2017-vn} applied local branching rates\cite{Neher2014-vl} and a more classical investigation of site-frequency spectra\cite{Fay2000-qw} to look for evidence of selective sweeps.

\paragraph*{Modeling lineage development}
The dynamic evolution of antibodies in germinal centers has been modeled for over a quarter century, for example leading to an early prediction of re-entry of circulating B cells back into the germinal center\cite{Kepler1993-xj}.
An early computer simulation framework, ``Clone,'' although not explicitly simulating an actual molecular sequence, simulated patterns of mutation in various parts of the BCR and their consequences\cite{Shlomchik1998-vb}.
Others have performed ABC-like (see first section for an introduction to ABC) analyses where they fit values such as mutation rate, selection, and clone affinity based on concordance of summary statistics\cite{Kleinstein2003-mu,Magori-Cohen2006-wd,Shahaf2008-cc}.
These analyses have typically been independent of existing population genetics theory, although recent work\cite{Horns2017-vn} makes use of site-frequency spectrum tools from population genetics.
Another vein of work uses agent-based and differential equation-based modeling to iteratively improve compartmental models of B cell development\cite{Kim2009-dr,Kleinstein2001-mk,Mehr2003-jz,Shahaf2004-kd,Shahaf2006-fn,Shahaf2010-na,Childs2015-cn,Wang2015-os,Wang2017-qu,Amitai2017-ku}.
For chronic infections such as HIV, antibody-pathogen coevolution certainly plays a role\cite{Liao2013-cr} although the dynamics between antibody emergence and viral escape are difficult to pin down\cite{Luo2015-qv}.
Some researchers have found a ``trunk-canopy'' tree structure from mature sequence data, in which a long ``trunk'' branch from the root extends from the naive sequence, after which there is a ``canopy'' of diversification\cite{Yaari2015-ss}.
However, it has been pointed out that the extent to which this structure is seen depends on the level of clustering\cite{Hershberg2015-hp}.

\optimist\
The previous review shows the disjointed state of the field: although clonal clustering, phylogenetics, selection inference, and modeling are all describing aspects of the same underlying process, they are divided into different problems (note that rearrangement inference, which is closely tied in with phylogenetic estimation, was relegated to its own section above, while isotype, which is closely tied with mutation processes on trees, appears in the next section!).
We must work towards unifying these various aspects in a shared framework.

Bayesian statistics offers a coherent framework for such information sharing and integration over uncertain latent states.
Although estimation of these complex posteriors will not be easy, we will be rewarded by more accurate inferences, leading to a more complete understanding of how affinity maturation works.
Our group is currently building on prior work\cite{Kepler2013-sy} to develop a Bayesian sampling procedure on trees that integrates out uncertainty in the unmutated common ancestor using a hidden Markov model.

We are also inspired by the work of Jonathan Laserson and colleagues\cite{Laserson2012-pi,Sok2013-td} who describes how sampling ancestral sequences explicitly as part of an MCMC can actually increase efficiency.
This echoes earlier work in a more general setting\cite{De_Koning2010-zc}.
The value of such sampling will be even greater when using more complex context-sensitive models, for which calculating likelihoods currently requires more intensive extensions to Gibbs sampling procedures (e.g., that of Chib\cite{Chib1995-ry}) to compute the marginal likelihood.
Other types of analysis, such as that of selection pressure\cite{McCoy2015-qi}, also require ancestral sequence inference.
Thus we believe that the next generation of phylogenetic algorithms for BCR sequences will infer a joint posterior on ancestral sequences and trees.

We also believe that tree-valued stochastic models will provide a unified foundation for learning about the diversification process from B cell sequence data.
Researchers working on viral populations have developed sophisticated tools for learning about viral spread by estimating ancestral population size using Bayesian ``skyline'' analysis\cite{Drummond2005-ks} and phylogenetic generalized linear models\cite{Dudas2017-sb}.
Somewhat analogous stochastic models for B cell development have been devised, but have only been used to generate distributions of summary statistics rather than being used for inference\cite{Magori-Cohen2006-wd}.
A more powerful tactic will be to develop models with parameters of interest and perform parameter inference directly.

Although repertoire-scale inference with a ``dream'' algorithm getting posterior distributions of all relevant parameters will not be possible, we can scale our computational ambition to the question at hand.
If we are very interested in a specific clonal family, it may be worth expending considerable computational effort in order to get high quality inferences for that family.
This may include probabilistically sampling alternative clusterings of that clonal family.
On the other hand, if we are looking for repertoire-level characteristics we will want to scale back our effort on each individual family in order to get an overall picture (although it is important that such algorithms are unbiased).

\lunatic\
Realistic forward-time models are essential to help guide the design and implementation of inferential algorithms.
For example, there is currently a need for models of affinity maturation that generate nucleotide sequences and trees interdependently with some level of realism.
Although antibody affinity models are relatively old\cite{Perelson1979-fp} and plentiful (see above), we aren't aware of any that generate nucleotide sequences.
Our group is currently developing such a model as part of a benchmarking exercise of ancestral reconstruction methods.

\subsection*{B cell clonal expansion: isotype}
Antibodies have an isotype-determining constant region that establishes the function of the antibody in the immune system.
Isotype can change through class-switch recombination, which arises due to double-stranded breaks resulting from AID deamination\cite{Hwang2015-wt,Methot2017-gi}.
High-throughput sequencing including isotype information is now available, and is shining light on this process.
For example, there are significant differences between isotypes in terms of their levels of somatic hypermutation\cite{Jackson2014-mo}.
This new data is also elucidating the rate with which antibodies switch isotype classes\cite{Horns2016-nk,Looney2016-fc}.
An analysis of sister lineages on either side of a branch point has suggested that the probability of switching to the various other isotypes is determined by more than just the current isotype \cite{Horns2016-nk}: rather, there is some additional hidden factor that determines the switching probability.
This could be summarized by saying that the isotype-switching process does not satisfy the Markov property.

\optimist\
If we do assume the Markov property, one can formulate an isotype model using existing continuous-time trait models.
Inference under such models is well developed from both maximum-likelihood\cite{Pagel1994-ab} and Bayesian\cite{Pagel2004-ex} perspectives.
Adding isotype as a hidden state in phylogenetic inference would be straightforward.

\lunatic\
One may also wish to model a non-Markov latent state for which existing inferential techniques will not apply.
Another interesting type of model would be one in which mutation and isotype-switching are linked probabilistically.

\subsection*{Estimating the complete adaptive immune response}
Although repertoire sequencing offers a remarkable perspective into the complex process of immune state, it will always offer an incomplete picture of the immune response.
First, it is well-acknowledged that we are taking a small sample from a very large population.
As such, it is common to extrapolate the total number of unique immune receptors from a sample\cite{Kaplinsky2016-gt}.

However, this is not the whole story: the common practice of sequencing from blood may not reflect what is happening with B and T cells in other compartments.
Recent work is beginning to lay the foundation for understanding the whole B and T cell response from blood samples.
This has included a ``B cell atlas'' of samples from many tissues of organ donors\cite{Meng2017-im}, as well as sequential fine needle aspirates from rhesus\cite{Havenar-Daughton2016-vk}, and sequencing from individual germinal centers using lymph node dissection in mice\cite{Tas2016-lq}.

In another direction one would like to understand the essential role that circulating antibodies play in the immune response.
Although B cell sequencing gives some idea of what antibodies can be made, it is certainly not the same as assaying the antibodies present in an individual.
The soluble antibody repertoire is determined by expression and antibody lifetime.
To do this, recent work has combined protein mass spectrometry with antibody sequencing\cite{Wine2013-ux,Lavinder2014-tx}.
Hopefully new protein sequencing methods\cite{Nivala2013-ic} will expand our perspective on soluble antibodies.

One may continue along these lines and say that even the pool of circulating antibodies are not the most interesting factor, and rather one should be interested in the collection of antigens that can be bound by those antibodies.
For this, recent work has used antigen microarrays\cite{Doolan2008-xw,Hertz2017-gt} to infer what peptides can be bound by circulating antibodies.
For T cells, yeast display has been used to identify the peptide specificity of TCRs found in cancer\cite{Gee2017-dt}.
Abstracting one notch further, one can use immunological assays between viral strains to assay an antigenic ``distance'' between them\cite{Smith2004-ub,Katzelnick2015-rr} that captures cross-reactivity of antibodies.

These complexities are well known to theoreticians.
The doctrine of ``original antigenic sin'' is over 60 years old\cite{Francis1960-nz} and modern methods continue to support past exposures as being essential for future development \cite{Andrews2015-ls,De_Bourcy2017-ki,Horwitz2017-cz}.
Perhaps the closest analysis of actual sequences are models of population-level immunity in which the fitness of a given influenza sequence is in part determined by its similarity to existing sequences to which the population is already presumably immune\cite{Luksza2014-xf,Neher2016-mn}.
There are also controlled experiments and mathematical models working to understand the impact of antibody feedback\cite{Zhang2013-yn,Childs2015-cn,Wang2015-os,Zarnitsyna2015-cn,Zarnitsyna2016-ev}, although this work hasn't been generalized to an inferential framework that can be used to understand individual repertoire data sets.
``Mutational antigenic profiling'' \cite{Doud2017-fj,Dingens2017-ar,Doud2018-oi}, which reveals how mutating an antigen can change antibody binding, and ``deep mutational scanning'', used to understand the impact of antibody sequence variation on binding\cite{Forsyth2013-zh,Adams2016-ja}, may be helpful in these efforts.

\optimist\
Given a lot of ``B cell atlas'' type data, one might be able to develop a migration model between the various compartments and infer rates based on observations of the same or related clones in different compartments.
The challenge with such a project will be to untangle re-seeding from early seeding and partial persistence.
Also, using such data, one may be able to model cell population sizes of difficult-to-sample compartments from ones that are easier to sample.

\lunatic\
One could dream of a model that attempts to capture the antigenic space that is covered by existing circulating antibodies.
In particular, recent efforts introduce antibody landscapes in the context of influenza \cite{Fonville2014-rl}.

\subsection*{Conclusion}
We have reviewed opportunities for probabilistic modeling in B and T cell sequence analysis.
To summarize, probabilistic models have a lot to contribute to rearrangement and lineage inference.
However, inferences on the functional properties of individual sequences (such as for initial selective filters or antigen binding) seem better done with machine learning methods rather than generative models for which likelihood calculation is not tractable.
For those aspects well suited to probabilistic modeling, we will be rewarded for integrating various aspects into a single framework where one level can communicate important information, including uncertainty, to another level.
For example, from B cell sequence analysis
\begin{itemize}
\item There is considerable signal in patterns of shared mutation that can help guide clustering inference, and the correct way of doing so is to combine phylogenetic inference with clustering.
\item It is common to include an inferred, unmutated common ancestor into a sequence alignment for phylogenetic inference.
Accounting for the corresponding uncertainty is important to gain accurate inferences on the processes that led to the observed sequences.
\item Phylogenetic trees are also uncertain, and disregarding that uncertainty will skew our downstream analyses of selection and models.
\end{itemize}
This model hierarchy can extend beyond the single-sample level to individual-level analysis through time, or population-level analysis.
The parameters we learn from these larger studies, such as germline gene existence and frequency, can feed back down to improve per-sample analysis.
They can also be used to analyze predictors of individual-level immune variation\cite{Brodin2017-qq}.

The computational statistician interested in immune receptor modeling is blessed with a complex biological system to analyze, intractable computational problems heaped on top of one another, and an ever-expanding collection of data sets generated from various in-vivo and in-vitro perturbations.
New methods are needed to perform inference under complex hierarchical models of immune receptor development for the optimistic program laid out in this paper to become a reality.
Although the field of computational immunology dates back many decades, we can gain inspiration and adapt techniques from the even longer tradition of macroevolutionary and ecological theory.
There, we have seen a complex interplay of generative models, summary statistics, and inferential models that have enabled the field's progress.

\section*{Acknowledgements}
We thank the AIRR community (\url{http://airr-community.org}) for discussions that have greatly enriched our understanding and appreciation of the field.
Christian Busse, Kristian Davidsen, William DeWitt, Julia Fukuyama, David Shaw, Duncan Ralph, and Corey Watson provided helpful feedback on this manuscript.
We thank Curt Callan for allowing us to use parts of his figure in our Figure~\ref{fig:probabilistic}.

This work supported by National Institutes of Health grants R01 GM113246, R01 AI12096, and U19 AI117891.
The research of Frederick Matsen was supported in part by a Faculty Scholar grant from the Howard Hughes Medical Institute and the Simons Foundation.


\bibliographystyle{vancouver}
\bibliography{main}

\begin{thebibliography}{100}

\bibitem{Sompayrac2011-fk}
Sompayrac LM.
\newblock How the immune system works.
\newblock Hoboken, New Jersey: John Wiley \& Sons; 2011.

\bibitem{Robins2013-ww}
Robins H.
\newblock Immunosequencing: applications of immune repertoire deep sequencing.
\newblock Curr Opin Immunol. 2013 Oct;25(5):646--652.
\newblock Available from: \url{http://dx.doi.org/10.1016/j.coi.2013.09.017}.

\bibitem{Yaari2015-pc}
Yaari G, Kleinstein SH.
\newblock Practical guidelines for B-cell receptor repertoire sequencing
  analysis.
\newblock Genome Med. 2015;7(1):121.
\newblock Available from: \url{http://dx.doi.org/10.1186/s13073-015-0243-2}.

\bibitem{Murugan2012-ue}
Murugan A, Mora T, Walczak AM, Callan CG Jr.
\newblock Statistical inference of the generation probability of T-cell
  receptors from sequence repertoires.
\newblock Proc Natl Acad Sci U S A. 2012 Oct;109(40):16161--16166.
\newblock Available from: \url{http://dx.doi.org/10.1073/pnas.1212755109}.

\bibitem{Elhanati2015-ld}
Elhanati Y, Sethna Z, Marcou Q, Callan CG Jr, Mora T, Walczak AM.
\newblock Inferring processes underlying B-cell repertoire diversity.
\newblock Philos Trans R Soc Lond B Biol Sci. 2015 Sep;370(1676).
\newblock Available from: \url{http://dx.doi.org/10.1098/rstb.2014.0243}.

\bibitem{Ralph2016-kr}
Ralph DK, Matsen FA IV.
\newblock Consistency of {VDJ} Rearrangement and Substitution Parameters
  Enables Accurate {B} Cell Receptor Sequence Annotation.
\newblock PLoS Comput Biol. 2016 Jan;12(1):e1004409.
\newblock Available from: \url{http://dx.doi.org/10.1371/journal.pcbi.1004409}.

\bibitem{Ralph2016-yl}
Ralph DK, Matsen FA IV.
\newblock {Likelihood-Based} Inference of {B} Cell Clonal Families.
\newblock PLoS Comput Biol. 2016 Oct;12(10):e1005086.
\newblock Available from:
  \url{http://journals.plos.org/ploscompbiol/article/asset?id=10.1371/journal.pcbi.1005086.PDF}.

\bibitem{MacKay2003-dm}
MacKay DJ.
\newblock Information Theory, Inference, and Learning Algorithms.
\newblock Cambridge, UK: Cambridge University Press; 2003.
\newblock Available from: \url{http://www.inference.org.uk/itprnn/book.pdf}.

\bibitem{Hornik2003-fu}
Hornik K, Leisch F, Zeileis A.
\newblock {JAGS}: A program for analysis of Bayesian graphical models using
  Gibbs sampling.
\newblock In: Proceedings of {DSC}. vol.~2. r-project.org; 2003. p. 1--1.
\newblock Available from:
  \url{https://www.r-project.org/conferences/DSC-2003/Proceedings/Plummer.pdf}.

\bibitem{Lunn2009-sv}
Lunn D, Spiegelhalter D, Thomas A, Best N.
\newblock The {BUGS} project: Evolution, critique and future directions.
\newblock Stat Med. 2009 Nov;28(25):3049--3067.
\newblock Available from: \url{http://dx.doi.org/10.1002/sim.3680}.

\bibitem{Carpenter2016-dn}
Carpenter B, Gelman A, Hoffman M, Lee D, Goodrich B, Betancourt M, et~al.
\newblock Stan: A probabilistic programming language.
\newblock J Stat Softw. 2016;20(2):1--37.
\newblock Available from:
  \url{http://datascienceassn.org/sites/default/files/Stan%20-%20Probabilistic%20Programming%20Language.pdf}.

\bibitem{Nascimento2017-wf}
Nascimento FF, dos Reis M, Yang Z.
\newblock A biologist’s guide to Bayesian phylogenetic analysis.
\newblock Nature Ecology \& Evolution. 2017 Sep;1(10):1446.
\newblock Available from:
  \url{https://www.nature.com/articles/s41559-017-0280-x}.

\bibitem{Drummond2012-mh}
Drummond AJ, Suchard MA, Xie D, Rambaut A.
\newblock Bayesian phylogenetics with {BEAUti} and the {BEAST} 1.7.
\newblock Mol Biol Evol. 2012;29(8):1969--1973.
\newblock Available from:
  \url{https://academic.oup.com/mbe/article-abstract/29/8/1969/1044583}.

\bibitem{Ronquist2012-hi}
Ronquist F, Teslenko M, van~der Mark P, Ayres DL, Darling A, H{\"o}hna S,
  et~al.
\newblock {MrBayes} 3.2: efficient Bayesian phylogenetic inference and model
  choice across a large model space.
\newblock Syst Biol. 2012 Feb;61(3):539--542.
\newblock Available from: \url{http://dx.doi.org/10.1093/sysbio/sys029}.

\bibitem{Dudas2017-sb}
Dudas G, Carvalho LM, Bedford T, Tatem AJ, Baele G, Faria NR, et~al.
\newblock Virus genomes reveal factors that spread and sustained the Ebola
  epidemic.
\newblock Nature. 2017 Apr;Available from:
  \url{http://dx.doi.org/10.1038/nature22040}.

\bibitem{Efron2012-go}
Efron B.
\newblock {Large-Scale} Inference: Empirical Bayes Methods for Estimation,
  Testing, and Prediction.
\newblock Cambridge University Press; 2012.
\newblock Available from:
  \url{https://market.android.com/details?id=book-kfE7VkqbGawC}.

\bibitem{Felsenstein1981-zs}
Felsenstein J.
\newblock Evolutionary trees from {DNA} sequences: a maximum likelihood
  approach.
\newblock J Mol Evol. 1981;17(6):368--376.
\newblock Available from: \url{http://www.ncbi.nlm.nih.gov/pubmed/7288891}.

\bibitem{Roch2006-ub}
Roch S.
\newblock A short proof that phylogenetic tree reconstruction by maximum
  likelihood is hard.
\newblock IEEE/ACM Trans Comput Biol Bioinform. 2006 Jan;3(1):92--94.
\newblock Available from: \url{http://dx.doi.org/10.1109/TCBB.2006.4}.

\bibitem{Durbin1998-uq}
Durbin R, Eddy SR, Krogh A, Mitchison G.
\newblock Biological Sequence Analysis: Probabilistic Models of Proteins and
  Nucleic Acids.
\newblock Cambridge: Cambridge University Press; 1998.
\newblock Available from: \url{http://books.google.com/books?id=R5P2GlJvigQC}.

\bibitem{Scott2011-zf}
Scott SL.
\newblock Bayesian methods for hidden Markov models.
\newblock J Am Stat Assoc. 2011;Available from:
  \url{http://amstat.tandfonline.com/doi/abs/10.1198/016214502753479464}.

\bibitem{Rogozin1992-xv}
Rogozin IB, Kolchanov NA.
\newblock Somatic hypermutagenesis in immunoglobulin genes. {II}. Influence of
  neighbouring base sequences on mutagenesis.
\newblock Biochim Biophys Acta. 1992 Nov;1171(1):11--18.
\newblock Available from: \url{http://www.ncbi.nlm.nih.gov/pubmed/1420357}.

\bibitem{Yaari2013-dg}
Yaari G, Vander~Heiden JA, Uduman M, Gadala-Maria D, Gupta N, Stern JNH, et~al.
\newblock Models of somatic hypermutation targeting and substitution based on
  synonymous mutations from high-throughput immunoglobulin sequencing data.
\newblock Front Immunol. 2013 Nov;4:358.
\newblock Available from: \url{http://dx.doi.org/10.3389/fimmu.2013.00358}.

\bibitem{Wei1990-yq}
Wei GCG, Tanner MA.
\newblock A Monte Carlo Implementation of the {EM} Algorithm and the Poor Man's
  Data Augmentation Algorithms.
\newblock J Am Stat Assoc. 1990;85(411):699--704.
\newblock Available from:
  \url{http://www.tandfonline.com/doi/abs/10.1080/01621459.1990.10474930}.

\bibitem{Feng2017-dh}
Feng J, Shaw DA, Minin VN, Simon N, Matsen FA IV.
\newblock Survival analysis of {DNA} mutation motifs with penalized
  proportional hazards. 2017 Nov;Available from:
  \url{http://arxiv.org/abs/1711.04057}.

\bibitem{Marin2012-gt}
Marin JM, Pudlo P, Robert CP, Ryder RJ.
\newblock Approximate Bayesian computational methods.
\newblock Stat Comput. 2012 Nov;22(6):1167--1180.
\newblock Available from:
  \url{https://link.springer.com/article/10.1007/s11222-011-9288-2}.

\bibitem{Robert2017-ki}
Robert PA, Rastogi A, Binder SC, Meyer-Hermann M.
\newblock How to Simulate a Germinal Center.
\newblock In: Calado DP, editor. Germinal Centers: Methods and Protocols.
  Methods in Molecular Biology. New York, NY: Springer New York; 2017. p.
  303--334.
\newblock Available from: \url{https://doi.org/10.1007/978-1-4939-7095-7_22}.

\bibitem{Kleinstein2003-mu}
Kleinstein SH, Louzoun Y, Shlomchik MJ.
\newblock Estimating hypermutation rates from clonal tree data.
\newblock J Immunol. 2003 Nov;171(9):4639--4649.
\newblock Available from: \url{https://www.ncbi.nlm.nih.gov/pubmed/14568938}.

\bibitem{Magori-Cohen2006-wd}
Magori-Cohen R, Louzoun Y, Kleinstein SH.
\newblock Mutation parameters from {DNA} sequence data using graph theoretic
  measures on lineage trees.
\newblock Bioinformatics. 2006 Jul;22(14):e332--40.
\newblock Available from:
  \url{http://dx.doi.org/10.1093/bioinformatics/btl239}.

\bibitem{Shahaf2008-cc}
Shahaf G, Barak M, Zuckerman NS, Swerdlin N, Gorfine M, Mehr R.
\newblock Antigen-driven selection in germinal centers as reflected by the
  shape characteristics of immunoglobulin gene lineage trees: a large-scale
  simulation study.
\newblock J Theor Biol. 2008 13~Aug;255(2):210--222.
\newblock Available from: \url{http://dx.doi.org/10.1016/j.jtbi.2008.08.005}.

\bibitem{Kepler1993-xj}
Kepler TB, Perelson AS.
\newblock Cyclic re-entry of germinal center {B} cells and the efficiency of
  affinity maturation.
\newblock Immunol Today. 1993 Aug;14(8):412--415.
\newblock Available from: \url{http://dx.doi.org/10.1016/0167-5699(93)90145-B}.

\bibitem{Victora2010-zg}
Victora GD, Schwickert TA, Fooksman DR, Kamphorst AO, Meyer-Hermann M, Dustin
  ML, et~al.
\newblock Germinal center dynamics revealed by multiphoton microscopy with a
  photoactivatable fluorescent reporter.
\newblock Cell. 2010 Nov;143(4):592--605.
\newblock Available from: \url{http://dx.doi.org/10.1016/j.cell.2010.10.032}.

\bibitem{Mesin2016-ew}
Mesin L, Ersching J, Victora GD.
\newblock Germinal Center {B} Cell Dynamics.
\newblock Immunity. 2016 Sep;45(3):471--482.
\newblock Available from: \url{http://dx.doi.org/10.1016/j.immuni.2016.09.001}.

\bibitem{Blei2014-wn}
Blei DM.
\newblock Build, Compute, Critique, Repeat: Data Analysis with Latent Variable
  Models.
\newblock Annu Rev Stat Appl. 2014 Jan;1(1):203--232.
\newblock Available from:
  \url{https://doi.org/10.1146/annurev-statistics-022513-115657}.

\bibitem{Bassing2002-bf}
Bassing CH, Swat W, Alt FW.
\newblock The mechanism and regulation of chromosomal {V(D)J} recombination.
\newblock Cell. 2002 Apr;109 Suppl:S45--55.
\newblock Available from: \url{http://www.ncbi.nlm.nih.gov/pubmed/11983152}.

\bibitem{Watson2012-ea}
Watson CT, Breden F.
\newblock The immunoglobulin heavy chain locus: genetic variation, missing
  data, and implications for human disease.
\newblock Genes Immun. 2012 Jul;13(5):363--373.
\newblock Available from: \url{http://dx.doi.org/10.1038/gene.2012.12}.

\bibitem{Jackson2013-ad}
Jackson KJL, Kidd MJ, Wang Y, Collins AM.
\newblock The Shape of the Lymphocyte Receptor Repertoire: Lessons from the {B}
  Cell Receptor.
\newblock Front Immunol. 2013 Sep;4:263.
\newblock Available from: \url{http://dx.doi.org/10.3389/fimmu.2013.00263}.

\bibitem{Rowen1996-pb}
Rowen L, Koop BF, Hood L.
\newblock The complete 685-kilobase {DNA} sequence of the human beta {T} cell
  receptor locus.
\newblock Science. 1996 Jun;272(5269):1755--1762.
\newblock Available from: \url{https://www.ncbi.nlm.nih.gov/pubmed/8650574}.

\bibitem{Boysen1997-bp}
Boysen C, Simon MI, Hood Genome~research L, {1997}.
\newblock Analysis of the 1.1-Mb human $\alpha$/$\delta$ T-cell receptor locus
  with bacterial artificial chromosome clones.
\newblock genomecshlporg. 1997;Available from:
  \url{http://genome.cshlp.org/content/7/4/330.short}.

\bibitem{Watson2013-rm}
Watson CT, Steinberg KM, Huddleston J, Warren RL, Malig M, Schein J, et~al.
\newblock Complete Haplotype Sequence of the Human Immunoglobulin {Heavy-Chain}
  Variable, Diversity, and Joining Genes and Characterization of Allelic and
  {Copy-Number} Variation.
\newblock Am J Hum Genet. 2013 Apr;92(4):530--546.
\newblock Available from:
  \url{http://www.sciencedirect.com/science/article/pii/S0002929713001122}.

\bibitem{Watson2014-op}
Watson CT, Steinberg KM, Graves TA, Warren RL, Malig M, Schein J, et~al.
\newblock Sequencing of the human {IG} light chain loci from a hydatidiform
  mole {BAC} library reveals locus-specific signatures of genetic diversity.
\newblock Genes Immun. 2014 Oct;Available from:
  \url{http://dx.doi.org/10.1038/gene.2014.56}.

\bibitem{Feeney1996-ea}
Feeney AJ, Atkinson MJ, Cowan MJ, Escuro G, Lugo G.
\newblock A defective Vkappa {A2} allele in Navajos which may play a role in
  increased susceptibility to haemophilus influenzae type b disease.
\newblock J Clin Invest. 1996 May;97(10):2277--2282.
\newblock Available from: \url{http://dx.doi.org/10.1172/JCI118669}.

\bibitem{Wang2011-mk}
Wang Y, Jackson KJ, G{\"a}eta B, Pomat W, Siba P, Sewell WA, et~al.
\newblock Genomic screening by 454 pyrosequencing identifies a new human {IGHV}
  gene and sixteen other new {IGHV} allelic variants.
\newblock Immunogenetics. 2011 May;63(5):259--265.
\newblock Available from: \url{http://dx.doi.org/10.1007/s00251-010-0510-8}.

\bibitem{Gadala-Maria2015-co}
Gadala-Maria D, Yaari G, Uduman M, Kleinstein SH.
\newblock Automated analysis of high-throughput B-cell sequencing data reveals
  a high frequency of novel immunoglobulin {V} gene segment alleles.
\newblock Proceedings of the National Academy of Sciences. 2015 Feb;Available
  from: \url{http://www.pnas.org/content/early/2015/02/05/1417683112.abstract}.

\bibitem{Scheepers2015-gs}
Scheepers C, Shrestha RK, Lambson BE, Jackson KJL, Wright IA, Naicker D, et~al.
\newblock Ability to develop broadly neutralizing {HIV-1} antibodies is not
  restricted by the germline Ig gene repertoire.
\newblock J Immunol. 2015 May;194(9):4371--4378.
\newblock Available from: \url{http://dx.doi.org/10.4049/jimmunol.1500118}.

\bibitem{Watson2017-mr}
Watson CT, Glanville J, Marasco WA.
\newblock The Individual and Population Genetics of Antibody Immunity.
\newblock Trends Immunol. 2017 Jul;38(7):459--470.
\newblock Available from: \url{http://dx.doi.org/10.1016/j.it.2017.04.003}.

\bibitem{Avnir2016-vs}
Avnir Y, Watson CT, Glanville J, Peterson EC, Tallarico AS, Bennett AS, et~al.
\newblock {IGHV1-69} polymorphism modulates anti-influenza antibody
  repertoires, correlates with {IGHV} utilization shifts and varies by
  ethnicity.
\newblock Sci Rep. 2016 Feb;6:20842.
\newblock Available from: \url{http://dx.doi.org/10.1038/srep20842}.

\bibitem{Boyd2010-yj}
Boyd SD, Ga{\"e}ta BA, Jackson KJ, Fire AZ, Marshall EL, Merker JD, et~al.
\newblock Individual variation in the germline Ig gene repertoire inferred from
  variable region gene rearrangements.
\newblock J Immunol. 2010 Jun;184(12):6986--6992.

\bibitem{Kidd2012-bt}
Kidd MJ, Chen Z, Wang Y, Jackson KJ, Zhang L, Boyd SD, et~al.
\newblock The inference of phased haplotypes for the immunoglobulin {H} chain
  {V} region gene loci by analysis of {VDJ} gene rearrangements.
\newblock J Immunol. 2012 Feb;188(3):1333--1340.
\newblock Available from: \url{http://dx.doi.org/10.4049/jimmunol.1102097}.

\bibitem{Corcoran2016-yy}
Corcoran MM, Phad GE, N{\'e}stor VB, Stahl-Hennig C, Sumida N, Persson MAA,
  et~al.
\newblock Production of individualized {V} gene databases reveals high levels
  of immunoglobulin genetic diversity.
\newblock Nat Commun. 2016 Dec;7:13642.
\newblock Available from: \url{http://dx.doi.org/10.1038/ncomms13642}.

\bibitem{Ralph2017-ih}
Ralph DK, Matsen FA IV.
\newblock Per-sample immunoglobulin germline inference from {B} cell receptor
  deep sequencing data. 2017 Nov;Available from:
  \url{http://arxiv.org/abs/1711.05843}.

\bibitem{Kirik2017-qq}
Kirik U, Greiff L, Levander F, Ohlin M.
\newblock Parallel antibody germline gene and haplotype analyses support the
  validity of immunoglobulin germline gene inference and discovery.
\newblock Mol Immunol. 2017 Jul;87:12--22.
\newblock Available from: \url{http://dx.doi.org/10.1016/j.molimm.2017.03.012}.

\bibitem{Reams2015-bo}
Reams AB, Roth JR.
\newblock Mechanisms of gene duplication and amplification.
\newblock Cold Spring Harb Perspect Biol. 2015 Feb;7(2):a016592.
\newblock Available from: \url{http://dx.doi.org/10.1101/cshperspect.a016592}.

\bibitem{Luo2017-gp}
Luo S, Yu JA, Li H, Song YS.
\newblock Worldwide genetic variation of the {IGHV} and {TRBV} immune receptor
  gene families in humans.
\newblock bioRxiv. 2017 Jun;p. 155440.
\newblock Available from:
  \url{http://biorxiv.org/content/early/2017/06/26/155440}.

\bibitem{Luo2016-kg}
Luo S, Yu JA, Song YS.
\newblock Estimating Copy Number and Allelic Variation at the Immunoglobulin
  Heavy Chain Locus Using Short Reads.
\newblock PLoS Comput Biol. 2016 Sep;12(9):e1005117.
\newblock Available from:
  \url{http://journals.plos.org/ploscompbiol/article/asset?id=10.1371/journal.pcbi.1005117.PDF}.

\bibitem{Schatz2011-cm}
Schatz DG, Ji Y.
\newblock Recombination centres and the orchestration of {V} (D) {J}
  recombination.
\newblock Nat Rev Immunol. 2011;11(4):251--263.
\newblock Available from:
  \url{https://www.nature.com/nri/journal/v11/n4/abs/nri2941.html}.

\bibitem{Cowell2003-vh}
Cowell LG, Davila M, Yang K, Kepler TB, Kelsoe G.
\newblock Prospective estimation of recombination signal efficiency and
  identification of functional cryptic signals in the genome by statistical
  modeling.
\newblock J Exp Med. 2003 Jan;197(2):207--220.
\newblock Available from: \url{https://www.ncbi.nlm.nih.gov/pubmed/12538660}.

\bibitem{Montefiori2016-pw}
Montefiori L, Wuerffel R, Roqueiro D, Lajoie B, Guo C, Gerasimova T, et~al.
\newblock Extremely {Long-Range} Chromatin Loops Link Topological Domains to
  Facilitate a Diverse Antibody Repertoire.
\newblock Cell Rep. 2016 Feb;14(4):896--906.
\newblock Available from: \url{http://dx.doi.org/10.1016/j.celrep.2015.12.083}.

\bibitem{Kepler1996-kd}
Kepler TB, Borrero M, Rugerio B, McCray SK, Clarke SH.
\newblock Interdependence of {N} nucleotide addition and recombination site
  choice in {V(D)J} rearrangement.
\newblock The Journal of Immunology. 1996 Nov;157(10):4451--4457.
\newblock Available from:
  \url{http://www.jimmunol.org/content/157/10/4451.abstract}.

\bibitem{Volpe2008-oq}
Volpe JM, Kepler TB.
\newblock Large-scale analysis of human heavy chain {V(D)J} recombination
  patterns.
\newblock Immunome Res. 2008 Feb;4:3.
\newblock Available from: \url{http://dx.doi.org/10.1186/1745-7580-4-3}.

\bibitem{Elhanati2016-yq}
Elhanati Y, Marcou Q, Mora T, Walczak AM.
\newblock {repgenHMM}: a dynamic programming tool to infer the rules of immune
  receptor generation from sequence data.
\newblock Bioinformatics. 2016 Feb;Available from:
  \url{http://dx.doi.org/10.1093/bioinformatics/btw112}.

\bibitem{Sethna2017-yl}
Sethna Z, Elhanati Y, Dudgeon CS, Callan CG Jr, Levine AJ, Mora T, et~al.
\newblock Insights into immune system development and function from mouse
  T-cell repertoires.
\newblock Proc Natl Acad Sci U S A. 2017 Feb;Available from:
  \url{http://dx.doi.org/10.1073/pnas.1700241114}.

\bibitem{Wilson2000-bg}
Wilson PC, Wilson K, Liu YJ, Banchereau J, Pascual V, Capra JD.
\newblock Receptor revision of immunoglobulin heavy chain variable region genes
  in normal human {B} lymphocytes.
\newblock J Exp Med. 2000 Jun;191(11):1881--1894.
\newblock Available from: \url{http://www.ncbi.nlm.nih.gov/pubmed/10839804}.

\bibitem{Collins2004-hw}
Collins AM, Ikutani M, Puiu D, Buck GA, Nadkarni A, Gaeta B.
\newblock Partitioning of Rearranged Ig Genes by Mutation Analysis Demonstrates
  {D-D} Fusion and {V} Gene Replacement in the Expressed Human Repertoire.
\newblock The Journal of Immunology. 2004 Jan;172(1):340--348.
\newblock Available from: \url{http://www.jimmunol.org/content/172/1/340}.

\bibitem{Meng2014-zw}
Meng W, Jayaraman S, Zhang B, Schwartz GW, Daber RD, Hershberg U, et~al.
\newblock Trials and Tribulations with {VH} Replacement.
\newblock Front Immunol. 2014 Jan;5:10.
\newblock Available from: \url{http://dx.doi.org/10.3389/fimmu.2014.00010}.

\bibitem{Ohm-Laursen2006-yq}
Ohm-Laursen L, Nielsen M, Larsen SR, Barington T.
\newblock No evidence for the use of {DIR}, {D--D} fusions, chromosome 15 open
  reading frames or {VHreplacement} in the peripheral repertoire was found on
  application of an improved algorithm, {JointML}, to 6329 human immunoglobulin
  {H} rearrangements.
\newblock Immunology. 2006 Oct;119(2):265--277.
\newblock Available from:
  \url{http://dx.doi.org/10.1111/j.1365-2567.2006.02431.x}.

\bibitem{Lee2016-tt}
Lee DW, Khavrutskii I, Wallqvist A, Bavari S, Cooper CL, Chaudhury S.
\newblock {BRILIA}: Integrated tool for high-throughput annotation and lineage
  tree assembly of B-cell repertoires.
\newblock Front Immunol. 2016 Dec;7.
\newblock Available from:
  \url{http://journal.frontiersin.org/article/10.3389/fimmu.2016.00681/abstract}.

\bibitem{Gaeta2007-mz}
Ga{\"e}ta BA, Malming HR, Jackson KJL, Bain ME, Wilson P, Collins AM.
\newblock {iHMMune-align}: hidden Markov model-based alignment and
  identification of germline genes in rearranged immunoglobulin gene sequences.
\newblock Bioinformatics. 2007 Apr;23(13):1580--1587.
\newblock Available from:
  \url{http://dx.doi.org/10.1093/bioinformatics/btm147}.

\bibitem{Munshaw2010-mj}
Munshaw S, Kepler TB.
\newblock {SoDA2}: a Hidden Markov Model approach for identification of
  immunoglobulin rearrangements.
\newblock Bioinformatics. 2010 Apr;26(7):867--872.
\newblock Available from:
  \url{http://dx.doi.org/10.1093/bioinformatics/btq056}.

\bibitem{Marcou2018-sm}
Marcou Q, Mora T, Walczak AM.
\newblock High-throughput immune repertoire analysis with {IGoR}.
\newblock Nat Commun. 2018 Feb;9(1):561.
\newblock Available from: \url{http://dx.doi.org/10.1038/s41467-018-02832-w}.

\bibitem{Kidd2015-vt}
Kidd MJ, Jackson KJL, Boyd SD, Collins AM.
\newblock {DJ} Pairing during {VDJ} Recombination Shows Positional Biases That
  Vary among Individuals with Differing {IGHD} Locus Immunogenotypes.
\newblock J Immunol. 2015 Dec;Available from:
  \url{http://dx.doi.org/10.4049/jimmunol.1501401}.

\bibitem{Nadel1995-rv}
Nadel B, Feeney AJ.
\newblock Influence of coding-end sequence on coding-end processing in {V(D)J}
  recombination.
\newblock J Immunol. 1995 Nov;155(9):4322--4329.
\newblock Available from: \url{https://www.ncbi.nlm.nih.gov/pubmed/7594591}.

\bibitem{Nadel1997-lv}
Nadel B, Feeney AJ.
\newblock Nucleotide deletion and {P} addition in {V(D)J} recombination: a
  determinant role of the coding-end sequence.
\newblock Mol Cell Biol. 1997 Jul;17(7):3768--3778.
\newblock Available from: \url{https://www.ncbi.nlm.nih.gov/pubmed/9199310}.

\bibitem{Larimore2012-lo}
Larimore K, McCormick MW, Robins HS, Greenberg PD.
\newblock Shaping of human germline {IgH} repertoires revealed by deep
  sequencing.
\newblock J Immunol. 2012 Aug;189(6):3221--3230.
\newblock Available from: \url{http://dx.doi.org/10.4049/jimmunol.1201303}.

\bibitem{Meng2011-ht}
Meng W, Yunk L, Wang LS, Maganty A, Xue E, Cohen PL, et~al.
\newblock Selection of individual {VH} genes occurs at the pro-B to pre-B cell
  transition.
\newblock J Immunol. 2011 Aug;187(4):1835--1844.
\newblock Available from: \url{http://dx.doi.org/10.4049/jimmunol.1100207}.

\bibitem{Benichou2013-ej}
Benichou J, Glanville J, Prak ETL, Azran R, Kuo TC, Pons J, et~al.
\newblock The restricted {DH} gene reading frame usage in the expressed human
  antibody repertoire is selected based upon its amino acid content.
\newblock J Immunol. 2013 Jun;190(11):5567--5577.
\newblock Available from: \url{http://dx.doi.org/10.4049/jimmunol.1201929}.

\bibitem{Elhanati2014-mf}
Elhanati Y, Murugan A, Callan CG Jr, Mora T, Walczak AM.
\newblock Quantifying selection in immune receptor repertoires.
\newblock Proc Natl Acad Sci U S A. 2014 Jun;Available from:
  \url{http://dx.doi.org/10.1073/pnas.1409572111}.

\bibitem{DeKosky2013-iz}
DeKosky BJ, Ippolito GC, Deschner RP, Lavinder JJ, Wine Y, Rawlings BM, et~al.
\newblock High-throughput sequencing of the paired human immunoglobulin heavy
  and light chain repertoire.
\newblock Nat Biotechnol. 2013 Feb;31(2):166--169.
\newblock Available from: \url{http://dx.doi.org/10.1038/nbt.2492}.

\bibitem{Howie2015-vp}
Howie B, Sherwood AM, Berkebile AD, Berka J, Emerson RO, Williamson DW, et~al.
\newblock High-throughput pairing of {T} cell receptor $\alpha$ and $\beta$
  sequences.
\newblock Sci Transl Med. 2015 Aug;7(301):301ra131.
\newblock Available from: \url{http://dx.doi.org/10.1126/scitranslmed.aac5624}.

\bibitem{Grigaityte2017-dp}
Grigaityte K, Carter JA, Goldfless SJ, Jeffery EW, Hause RJ, Jiang Y, et~al.
\newblock Single-cell sequencing reveals $\alpha$$\beta$ chain pairing shapes
  the {T} cell repertoire; 2017.
\newblock Available from:
  \url{https://www.biorxiv.org/content/early/2017/11/02/213462}.

\bibitem{Emerson2017-co}
Emerson RO, DeWitt WS, Vignali M, Gravley J, Hu JK, Osborne EJ, et~al.
\newblock Immunosequencing identifies signatures of cytomegalovirus exposure
  history and {HLA-mediated} effects on the {T} cell repertoire.
\newblock Nat Genet. 2017 Apr;Available from:
  \url{http://dx.doi.org/10.1038/ng.3822}.

\bibitem{Desponds2016-ja}
Desponds J, Mora T, Walczak AM.
\newblock Fluctuating fitness shapes the clone-size distribution of immune
  repertoires.
\newblock Proc Natl Acad Sci U S A. 2016 Jan;113(2):274--279.
\newblock Available from: \url{http://dx.doi.org/10.1073/pnas.1512977112}.

\bibitem{Desponds2017-kj}
Desponds J, Mayer A, Mora T, Walczak AM.
\newblock Population dynamics of immune repertoires. 2017 Mar;Available from:
  \url{http://arxiv.org/abs/1703.00226}.

\bibitem{Qi2016-or}
Qi Q, Cavanagh MM, Le~Saux S, NamKoong H, Kim C, Turgano E, et~al.
\newblock Diversification of the antigen-specific {T} cell receptor repertoire
  after varicella zoster vaccination.
\newblock Sci Transl Med. 2016 Mar;8(332):332ra46.
\newblock Available from: \url{http://dx.doi.org/10.1126/scitranslmed.aaf1725}.

\bibitem{Pogorelyy2017-gx}
Pogorelyy MV, Elhanati Y, Marcou Q, Sycheva AL, Komech EA, Nazarov VI, et~al.
\newblock Persisting fetal clonotypes influence the structure and overlap of
  adult human {T} cell receptor repertoires.
\newblock PLoS Comput Biol. 2017 Jul;13(7):e1005572.
\newblock Available from: \url{http://dx.doi.org/10.1371/journal.pcbi.1005572}.

\bibitem{Boyd2013-lt}
Boyd SD, Liu Y, Wang C, Martin V, Dunn-Walters DK.
\newblock Human lymphocyte repertoires in ageing.
\newblock Curr Opin Immunol. 2013 Aug;25(4):511--515.
\newblock Available from: \url{http://dx.doi.org/10.1016/j.coi.2013.07.007}.

\bibitem{Dash2017-fr}
Dash P, Fiore-Gartland AJ, Hertz T, Wang GC, Sharma S, Souquette A, et~al.
\newblock Quantifiable predictive features define epitope-specific {T} cell
  receptor repertoires.
\newblock Nature. 2017 Jun;Available from:
  \url{http://dx.doi.org/10.1038/nature22383}.

\bibitem{Yokota2017-rr}
Yokota R, Kaminaga Y, Kobayashi TJ.
\newblock Quantification of {Inter-Sample} Differences in {T-Cell} Receptor
  Repertoires Using {Sequence-Based} Information.
\newblock Front Immunol. 2017;8:1500.
\newblock Available from:
  \url{https://www.frontiersin.org/article/10.3389/fimmu.2017.01500}.

\bibitem{Glanville2017-uw}
Glanville J, Huang H, Nau A, Hatton O, Wagar LE, Rubelt F, et~al.
\newblock Identifying specificity groups in the {T} cell receptor repertoire.
\newblock Nature. 2017 Jun;Available from:
  \url{http://dx.doi.org/10.1038/nature22976}.

\bibitem{Sharon2016-ib}
Sharon E, Sibener LV, Battle A, Fraser HB, Garcia KC, Pritchard JK.
\newblock Genetic variation in {MHC} proteins is associated with {T} cell
  receptor expression biases.
\newblock Nat Genet. 2016 Aug;Available from:
  \url{http://dx.doi.org/10.1038/ng.3625}.

\bibitem{Thomas2014-jo}
Thomas N, Best K, Cinelli M, Reich-Zeliger S, Gal H, Shifrut E, et~al.
\newblock Tracking global changes induced in the {CD4} T-cell receptor
  repertoire by immunization with a complex antigen using short stretches of
  {CDR3} protein sequence.
\newblock Bioinformatics. 2014 Nov;30(22):3181--3188.
\newblock Available from:
  \url{http://dx.doi.org/10.1093/bioinformatics/btu523}.

\bibitem{Ostmeyer2017-rz}
Ostmeyer J, Christley S, Rounds WH, Toby I, Greenberg BM, Monson NL, et~al.
\newblock Statistical classifiers for diagnosing disease from immune
  repertoires: a case study using multiple sclerosis.
\newblock BMC Bioinformatics. 2017 Sep;18(1):401.
\newblock Available from: \url{http://dx.doi.org/10.1186/s12859-017-1814-6}.

\bibitem{Pogorelyy2018-ox}
Pogorelyy MV, Minervina AA, Chudakov DM, Mamedov IZ, Lebedev YB, Mora T, et~al.
\newblock Method for identification of condition-associated public antigen
  receptor sequences.
\newblock Elife. 2018 Mar;7.
\newblock Available from: \url{http://dx.doi.org/10.7554/eLife.33050}.

\bibitem{Shugay2017-nj}
Shugay M, Bagaev DV, Zvyagin IV, Vroomans RM, Crawford JC, Dolton G, et~al.
\newblock {VDJdb}: a curated database of T-cell receptor sequences with known
  antigen specificity.
\newblock Nucleic Acids Res. 2017 Sep;Available from:
  \url{https://academic.oup.com/nar/article/doi/10.1093/nar/gkx760/4101254/VDJdb-a-curated-database-of-T-cell-receptor}.

\bibitem{Birnbaum2014-xw}
Birnbaum ME, Mendoza JL, Sethi DK, Dong S, Glanville J, Dobbins J, et~al.
\newblock Deconstructing the {peptide-MHC} specificity of {T} cell recognition.
\newblock Cell. 2014 May;157(5):1073--1087.
\newblock Available from: \url{http://dx.doi.org/10.1016/j.cell.2014.03.047}.

\bibitem{Victora2012-lu}
Victora GD, Nussenzweig MC.
\newblock Germinal centers.
\newblock Annu Rev Immunol. 2012 Jan;30:429--457.
\newblock Available from:
  \url{http://dx.doi.org/10.1146/annurev-immunol-020711-075032}.

\bibitem{Manz2005-lq}
Manz RA, Hauser AE, Hiepe F, Radbruch A.
\newblock Maintenance of serum antibody levels.
\newblock Annu Rev Immunol. 2005;23:367--386.
\newblock Available from:
  \url{http://dx.doi.org/10.1146/annurev.immunol.23.021704.115723}.

\bibitem{Galson2015-ey}
Galson JD, Kelly DF, Truck J.
\newblock Identification of {Antigen-Specific} {B-Cell} Receptor Sequences from
  the Total {B-Cell} Repertoire.
\newblock Crit Rev Immunol. 2015;35(6):463--478.
\newblock Available from:
  \url{http://dx.doi.org/10.1615/CritRevImmunol.2016016462}.

\bibitem{Jackson2014-rx}
Jackson KJL, Liu Y, Roskin KM, Glanville J, Hoh RA, Seo K, et~al.
\newblock Human Responses to Influenza Vaccination Show Seroconversion
  Signatures and Convergent Antibody Rearrangements.
\newblock Cell Host Microbe. 2014 Jun;Available from:
  \url{http://dx.doi.org/10.1016/j.chom.2014.05.013}.

\bibitem{Laserson2014-yh}
Laserson U, Vigneault F, Gadala-Maria D, Yaari G, Uduman M, Vander~Heiden JA,
  et~al.
\newblock High-resolution antibody dynamics of vaccine-induced immune
  responses.
\newblock Proc Natl Acad Sci U S A. 2014 Mar;Available from:
  \url{http://dx.doi.org/10.1073/pnas.1323862111}.

\bibitem{Jiang2013-kj}
Jiang N, He J, Weinstein JA, Penland L, Sasaki S, He XS, et~al.
\newblock Lineage structure of the human antibody repertoire in response to
  influenza vaccination.
\newblock Sci Transl Med. 2013 Feb;5(171):171ra19.
\newblock Available from: \url{http://dx.doi.org/10.1126/scitranslmed.3004794}.

\bibitem{Martin2015-dx}
Martin V, Bryan~Wu YC, Kipling D, Dunn-Walters D.
\newblock Ageing of the B-cell repertoire.
\newblock Philos Trans R Soc Lond B Biol Sci. 2015 Sep;370(1676).
\newblock Available from: \url{http://dx.doi.org/10.1098/rstb.2014.0237}.

\bibitem{Galson2015-cp}
Galson JD, Tr{\"u}ck J, Fowler A, Clutterbuck EA, M{\"u}nz M, Cerundolo V,
  et~al.
\newblock Analysis of {B} Cell Repertoire Dynamics Following Hepatitis {B}
  Vaccination in Humans, and Enrichment of Vaccine-specific Antibody Sequences.
\newblock EBioMedicine. 2015 Dec;2(12):2070--2079.
\newblock Available from: \url{http://dx.doi.org/10.1016/j.ebiom.2015.11.034}.

\bibitem{Galson2015-qk}
Galson JD, Clutterbuck EA, Tr{\"u}ck J, Ramasamy MN, M{\"u}nz M, Fowler A,
  et~al.
\newblock {BCR} repertoire sequencing: different patterns of B-cell activation
  after two Meningococcal vaccines.
\newblock Immunol Cell Biol. 2015 May;Available from:
  \url{http://dx.doi.org/10.1038/icb.2015.57}.

\bibitem{Galson2016-iq}
Galson JD, Tr{\"u}ck J, Clutterbuck EA, Fowler A, Cerundolo V, Pollard AJ,
  et~al.
\newblock B-cell repertoire dynamics after sequential hepatitis {B} vaccination
  and evidence for cross-reactive B-cell activation.
\newblock Genome Med. 2016 Jun;8(1):68.
\newblock Available from: \url{http://dx.doi.org/10.1186/s13073-016-0322-z}.

\bibitem{Scheid2009-ot}
Scheid JF, Mouquet H, Feldhahn N, Seaman MS, Velinzon K, Pietzsch J, et~al.
\newblock Broad diversity of neutralizing antibodies isolated from memory {B}
  cells in {HIV-infected} individuals.
\newblock Nature. 2009 Apr;458(7238):636--640.
\newblock Available from: \url{http://dx.doi.org/10.1038/nature07930}.

\bibitem{Yoon2015-ec}
Yoon H, Macke J, West AP Jr, Foley B, Bjorkman PJ, Korber B, et~al.
\newblock {CATNAP}: a tool to compile, analyze and tally neutralizing antibody
  panels.
\newblock Nucleic Acids Res. 2015 Jul;43(W1):W213--9.
\newblock Available from: \url{http://dx.doi.org/10.1093/nar/gkv404}.

\bibitem{Asti2016-uy}
Asti L, Uguzzoni G, Marcatili P, Pagnani A.
\newblock {Maximum-Entropy} Models of Sequenced Immune Repertoires Predict
  {Antigen-Antibody} Affinity.
\newblock PLoS Comput Biol. 2016 Apr;12(4):e1004870.
\newblock Available from: \url{http://dx.doi.org/10.1371/journal.pcbi.1004870}.

\bibitem{Galson2015-yl}
Galson JD, Tr{\"u}ck J, Fowler A, M{\"u}nz M, Cerundolo V, Pollard AJ, et~al.
\newblock {In-Depth} Assessment of {Within-Individual} and {Inter-Individual}
  Variation in the {B} Cell Receptor Repertoire.
\newblock Front Immunol. 2015 Oct;6:531.
\newblock Available from: \url{http://dx.doi.org/10.3389/fimmu.2015.00531}.

\bibitem{Henry_Dunand2015-hf}
Henry~Dunand CJ, Wilson PC.
\newblock Restricted, canonical, stereotyped and convergent immunoglobulin
  responses.
\newblock Philos Trans R Soc Lond B Biol Sci. 2015 Sep;370(1676).
\newblock Available from: \url{http://dx.doi.org/10.1098/rstb.2014.0238}.

\bibitem{Truck2015-aj}
Tr{\"u}ck J, Ramasamy MN, Galson JD, Rance R, Parkhill J, Lunter G, et~al.
\newblock Identification of antigen-specific {B} cell receptor sequences using
  public repertoire analysis.
\newblock J Immunol. 2015 Jan;194(1):252--261.
\newblock Available from: \url{http://dx.doi.org/10.4049/jimmunol.1401405}.

\bibitem{Wang2015-on}
Wang C, Liu Y, Cavanagh MM, Le~Saux S, Qi Q, Roskin KM, et~al.
\newblock B-cell repertoire responses to varicella-zoster vaccination in human
  identical twins.
\newblock Proc Natl Acad Sci U S A. 2015 Jan;112(2):500--505.
\newblock Available from: \url{http://dx.doi.org/10.1073/pnas.1415875112}.

\bibitem{Greiff2017-hi}
Greiff V, Menzel U, Miho E, Weber C, Riedel R, Cook S, et~al.
\newblock Systems Analysis Reveals High Genetic and {Antigen-Driven}
  Predetermination of Antibody Repertoires throughout {B} Cell Development.
\newblock Cell Rep. 2017 May;19(7):1467--1478.
\newblock Available from:
  \url{http://www.cell.com/article/S221112471730565X/abstract}.

\bibitem{Collins2017-te}
Collins AM, Jackson KJL.
\newblock On being the right size: antibody repertoire formation in the mouse
  and human.
\newblock Immunogenetics. 2017 Dec;Available from:
  \url{http://dx.doi.org/10.1007/s00251-017-1049-8}.

\bibitem{DeWitt2016-yy}
DeWitt WS, Lindau P, Snyder TM, Sherwood AM, Vignali M, Carlson CS, et~al.
\newblock A Public Database of Memory and Naive {B-Cell} Receptor Sequences.
\newblock PLoS One. 2016 Aug;11(8):e0160853.
\newblock Available from: \url{http://dx.doi.org/10.1371/journal.pone.0160853}.

\bibitem{DeWitt2014-sz}
DeWitt W, Lindau P, Snyder T, Vignali M, Emerson R, Robins H.
\newblock Replicate immunosequencing as a robust probe of {B} cell repertoire
  diversity. 2014 Oct;Available from: \url{http://arxiv.org/abs/1410.0350}.

\bibitem{Methot2017-gi}
Methot SP, Di~Noia JM.
\newblock Chapter Two - Molecular Mechanisms of Somatic Hypermutation and Class
  Switch Recombination.
\newblock In: {Frederick W  Alt}, editor. Advances in Immunology. vol. 133.
  Academic Press; 2017. p. 37--87.
\newblock Available from:
  \url{http://www.sciencedirect.com/science/article/pii/S0065277616300530}.

\bibitem{Hwang2017-tt}
Hwang JK, Wang C, Du Z, Meyers RM, Kepler TB, Neuberg D, et~al.
\newblock Sequence intrinsic somatic mutation mechanisms contribute to affinity
  maturation of {VRC01-class} {HIV-1} broadly neutralizing antibodies.
\newblock Proc Natl Acad Sci U S A. 2017 Jul;Available from:
  \url{http://dx.doi.org/10.1073/pnas.1709203114}.

\bibitem{Dunn-Walters1998-ds}
Dunn-Walters DK, Dogan A, Boursier L, MacDonald CM, Spencer J.
\newblock Base-specific sequences that bias somatic hypermutation deduced by
  analysis of out-of-frame human {IgVH} genes.
\newblock J Immunol. 1998 Mar;160(5):2360--2364.
\newblock Available from: \url{https://www.ncbi.nlm.nih.gov/pubmed/9498777}.

\bibitem{Cowell2000-rq}
Cowell LG, Kepler TB.
\newblock The {Nucleotide-Replacement} Spectrum Under Somatic Hypermutation
  Exhibits Microsequence Dependence That Is {Strand-Symmetric} and Distinct
  from That Under Germline Mutation.
\newblock The Journal of Immunology. 2000 Feb;164(4):1971--1976.
\newblock Available from:
  \url{http://www.jimmunol.org/content/164/4/1971.short}.

\bibitem{Pham2003-jm}
Pham P, Bransteitter R, Petruska J, Goodman MF.
\newblock Processive {AID-catalysed} cytosine deamination on single-stranded
  {DNA} simulates somatic hypermutation.
\newblock Nature. 2003 Jul;424(6944):103--107.
\newblock Available from: \url{http://dx.doi.org/10.1038/nature01760}.

\bibitem{Cui2016-wz}
Cui A, Di~Niro R, Vander~Heiden JA, Briggs AW, Adams K, Gilbert T, et~al.
\newblock A Model of Somatic Hypermutation Targeting in Mice Based on
  {High-Throughput} Ig Sequencing Data.
\newblock J Immunol. 2016 Nov;197(9):3566--3574.
\newblock Available from: \url{http://dx.doi.org/10.4049/jimmunol.1502263}.

\bibitem{Cohen2011-rs}
Cohen RM, Kleinstein SH, Louzoun Y.
\newblock Somatic hypermutation targeting is influenced by location within the
  immunoglobulin {V} region.
\newblock Mol Immunol. 2011 Jul;48(12-13):1477--1483.
\newblock Available from: \url{http://dx.doi.org/10.1016/j.molimm.2011.04.002}.

\bibitem{Sheng2017-ib}
Sheng Z, Schramm CA, Kong R, {NISC Comparative Sequencing Program}, Mullikin
  JC, Mascola JR, et~al.
\newblock {Gene-Specific} Substitution Profiles Describe the Types and
  Frequencies of Amino Acid Changes during Antibody Somatic Hypermutation.
\newblock Front Immunol. 2017 May;8:537.
\newblock Available from: \url{http://dx.doi.org/10.3389/fimmu.2017.00537}.

\bibitem{Kirik2017-bc}
Kirik U, Persson H, Levander F, Greiff L, Ohlin M.
\newblock Antibody Heavy Chain Variable Domains of Different Germline Gene
  Origins Diversify through Different Paths.
\newblock Front Immunol. 2017;8:1433.
\newblock Available from:
  \url{https://www.frontiersin.org/article/10.3389/fimmu.2017.01433}.

\bibitem{Dhar2018-pi}
Dhar A, Davidsen K, Matsen FA IV, Minin VN.
\newblock Predicting {B} Cell Receptor Substitution Profiles Using Public
  Repertoire Data. 2018 Feb;Available from:
  \url{http://arxiv.org/abs/1802.06406}.

\bibitem{Dunn-Walters1998-ra}
Dunn-Walters DK, Spencer J.
\newblock Strong intrinsic biases towards mutation and conservation of bases in
  human {IgVH} genes during somatic hypermutation prevent statistical analysis
  of antigen selection.
\newblock Immunology. 1998 Nov;95(3):339--345.
\newblock Available from: \url{http://www.ncbi.nlm.nih.gov/pubmed/9824495}.

\bibitem{Yaari2012-kk}
Yaari G, Uduman M, Kleinstein SH.
\newblock Quantifying selection in high-throughput Immunoglobulin sequencing
  data sets.
\newblock Nucleic Acids Res. 2012 May;40(17):e134.
\newblock Available from: \url{http://dx.doi.org/10.1093/nar/gks457}.

\bibitem{McCoy2015-qi}
McCoy CO, Bedford T, Minin VN, Bradley P, Robins H, Matsen FA IV.
\newblock Quantifying evolutionary constraints on B-cell affinity maturation.
\newblock Philos Trans R Soc Lond B Biol Sci. 2015 Sep;370(1676).
\newblock Available from: \url{http://dx.doi.org/10.1098/rstb.2014.0244}.

\bibitem{Vieira2018-yy}
Vieira MC, Zinder D, Cobey S.
\newblock Selection and neutral mutations drive pervasive mutability losses in
  long-lived {anti-HIV} {B} cell lineages.
\newblock Mol Biol Evol. 2018 Feb;Available from:
  \url{https://academic.oup.com/mbe/advance-article/doi/10.1093/molbev/msy024/4904157?rss=1&utm_source=dlvr.it&utm_medium=twitter}.

\bibitem{Spencer1999-tl}
Spencer J, Dunn M, Dunn-Walters DK.
\newblock Characteristics of sequences around individual nucleotide
  substitutions in {IgVH} genes suggest different {GC} and {AT} mutators.
\newblock J Immunol. 1999 Jun;162(11):6596--6601.
\newblock Available from: \url{https://www.ncbi.nlm.nih.gov/pubmed/10352276}.

\bibitem{Rogozin2001-pw}
Rogozin IB, Pavlov YI, Bebenek K, Matsuda T, Kunkel TA.
\newblock Somatic mutation hotspots correlate with {DNA} polymerase $\eta$
  error spectrum.
\newblock Nat Immunol. 2001 Jun;2(6):530--536.
\newblock Available from: \url{http://dx.doi.org/10.1038/88732}.

\bibitem{Wilson2005-qa}
Wilson TM, Vaisman A, Martomo SA, Sullivan P, Lan L, Hanaoka F, et~al.
\newblock {MSH2--MSH6} stimulates {DNA} polymerase $\eta$, suggesting a role
  for A: {T} mutations in antibody genes.
\newblock J Exp Med. 2005;201(4):637--645.
\newblock Available from:
  \url{http://jem.rupress.org/content/201/4/637.abstract}.

\bibitem{Wang2010-lf}
Wang M, Rada C, Neuberger MS.
\newblock Altering the spectrum of immunoglobulin {V} gene somatic
  hypermutation by modifying the active site of {AID}.
\newblock J Exp Med. 2010 Jan;207(1):141--153.
\newblock Available from: \url{http://dx.doi.org/10.1084/jem.20092238}.

\bibitem{Mak2013-iu}
Mak CH, Pham P, Afif SA, Goodman MF.
\newblock A mathematical model for scanning and catalysis on single-stranded
  {DNA}, illustrated with activation-induced deoxycytidine deaminase.
\newblock J Biol Chem. 2013 Oct;288(41):29786--29795.
\newblock Available from: \url{http://dx.doi.org/10.1074/jbc.M113.506550}.

\bibitem{Chahwan2012-gp}
Chahwan R, Edelmann W, Scharff MD, Roa S.
\newblock {AIDing} antibody diversity by error-prone mismatch repair.
\newblock Semin Immunol. 2012 Aug;24(4):293--300.
\newblock Available from: \url{http://dx.doi.org/10.1016/j.smim.2012.05.005}.

\bibitem{Wilson1998-br}
Wilson PC, de~Bouteiller O, Liu YJ, Potter K, Banchereau J, Capra JD, et~al.
\newblock Somatic hypermutation introduces insertions and deletions into
  immunoglobulin {V} genes.
\newblock J Exp Med. 1998 Jan;187(1):59--70.
\newblock Available from: \url{https://www.ncbi.nlm.nih.gov/pubmed/9419211}.

\bibitem{Briney2012-sv}
Briney BS, Willis JR, Crowe JE Jr.
\newblock Location and length distribution of somatic hypermutation-associated
  {DNA} insertions and deletions reveals regions of antibody structural
  plasticity.
\newblock Genes Immun. 2012 Oct;13(7):523--529.
\newblock Available from: \url{http://dx.doi.org/10.1038/gene.2012.28}.

\bibitem{Bowers2014-no}
Bowers PM, Verdino P, Wang Z, da~Silva~Correia J, Chhoa M, Macondray G, et~al.
\newblock Nucleotide insertions and deletions complement point mutations to
  massively expand the diversity created by somatic hypermutation of
  antibodies.
\newblock J Biol Chem. 2014 Nov;289(48):33557--33567.
\newblock Available from: \url{http://dx.doi.org/10.1074/jbc.M114.607176}.

\bibitem{Yeap2015-nl}
Yeap LS, Hwang JK, Du Z, Meyers RM, Meng FL, Jakubauskait{\. e} A, et~al.
\newblock {Sequence-Intrinsic} Mechanisms that Target {AID} Mutational Outcomes
  on Antibody Genes.
\newblock Cell. 2015;Available from:
  \url{http://www.sciencedirect.com/science/article/pii/S0092867415013975}.

\bibitem{Kepler2014-pg}
Kepler TB, Liao HX, Alam SM, Bhaskarabhatla R, Zhang R, Yandava C, et~al.
\newblock Immunoglobulin Gene Insertions and Deletions in the Affinity
  Maturation of {HIV-1} Broadly Reactive Neutralizing Antibodies.
\newblock Cell Host Microbe. 2014 Sep;16(3):304--313.
\newblock Available from: \url{http://dx.doi.org/10.1016/j.chom.2014.08.006}.

\bibitem{Hwang2004-pj}
Hwang DG, Green P.
\newblock Bayesian Markov chain Monte Carlo sequence analysis reveals varying
  neutral substitution patterns in mammalian evolution.
\newblock Proceedings of the National Academy of Sciences USA. 2004
  28~Sep;101(39):13994--14001.
\newblock Available from: \url{http://dx.doi.org/10.1073/pnas.0404142101}.

\bibitem{Hobolth2008-dx}
Hobolth A.
\newblock A {Markov} chain {Monte Carlo} expectation maximization algorithm for
  statistical analysis of {DNA} sequence evolution with neighbor-dependent
  substitutio n rates.
\newblock Journal of Computational and Graphical Statistics.
  2008;17(1):138--162.
\newblock Available from: \url{http://dx.doi.org/10.1198/106186008X289010}.

\bibitem{Bishop1986-to}
Bishop MJ, Thompson EA.
\newblock Maximum likelihood alignment of {DNA} sequences.
\newblock J Mol Biol. 1986 Jul;190(2):159--165.
\newblock Available from: \url{https://www.ncbi.nlm.nih.gov/pubmed/3641921}.

\bibitem{Thorne1991-fm}
Thorne JL, Kishino H, Felsenstein J.
\newblock An evolutionary model for maximum likelihood alignment of {DNA}
  sequences.
\newblock J Mol Evol. 1991 Aug;33(2):114--124.
\newblock Available from: \url{https://www.ncbi.nlm.nih.gov/pubmed/1920447}.

\bibitem{Thorne1992-nk}
Thorne JL, Kishino H, Felsenstein J.
\newblock Inching toward reality: an improved likelihood model of sequence
  evolution.
\newblock J Mol Evol. 1992 Jan;34(1):3--16.
\newblock Available from: \url{https://www.ncbi.nlm.nih.gov/pubmed/1556741}.

\bibitem{Bouchard-Cote2013-ax}
Bouchard-C{\^o}t{\'e} A, Jordan MI.
\newblock Evolutionary inference via the Poisson Indel Process.
\newblock Proc Natl Acad Sci U S A. 2013 Jan;110(4):1160--1166.
\newblock Available from: \url{http://dx.doi.org/10.1073/pnas.1220450110}.

\bibitem{Zhai2017-ue}
Zhai Y, Alexandre BC.
\newblock A Poissonian Model of Indel Rate Variation for Phylogenetic Tree
  Inference.
\newblock Syst Biol. 2017 Sep;66(5):698--714.
\newblock Available from: \url{http://dx.doi.org/10.1093/sysbio/syx033}.

\bibitem{Mayer2017-ij}
Mayer CT, Gazumyan A, Kara EE, Gitlin AD, Golijanin J, Viant C, et~al.
\newblock The microanatomic segregation of selection by apoptosis in the
  germinal center.
\newblock Science. 2017 Sep;Available from:
  \url{http://dx.doi.org/10.1126/science.aao2602}.

\bibitem{Tas2016-lq}
Tas JMJ, Mesin L, Pasqual G, Targ S, Jacobsen JT, Mano YM, et~al.
\newblock Visualizing antibody affinity maturation in germinal centers.
\newblock Science. 2016 Mar;351(6277):1048--1054.
\newblock Available from: \url{http://dx.doi.org/10.1126/science.aad3439}.

\bibitem{Kuraoka2016-zs}
Kuraoka M, Schmidt AG, Nojima T, Feng F, Watanabe A, Kitamura D, et~al.
\newblock Complex Antigens Drive Permissive Clonal Selection in Germinal
  Centers.
\newblock Immunity. 2016 Mar;Available from:
  \url{http://dx.doi.org/10.1016/j.immuni.2016.02.010}.

\bibitem{Zhang2013-yn}
Zhang Y, Meyer-Hermann M, George LA, Figge MT, Khan M, Goodall M, et~al.
\newblock Germinal center {B} cells govern their own fate via antibody
  feedback.
\newblock J Exp Med. 2013 Feb;210(3):457--464.
\newblock Available from: \url{http://dx.doi.org/10.1084/jem.20120150}.

\bibitem{De_Bourcy2017-ki}
de~Bourcy CFA, Angel CJL, Vollmers C, Dekker CL, Davis MM, Quake SR.
\newblock Phylogenetic analysis of the human antibody repertoire reveals
  quantitative signatures of immune senescence and aging.
\newblock Proc Natl Acad Sci U S A. 2017 Jan;114(5):1105--1110.
\newblock Available from: \url{http://dx.doi.org/10.1073/pnas.1617959114}.

\bibitem{Or-Guil2007-vd}
Or-Guil M, Wittenbrink N, Weiser AA, Schuchhardt J.
\newblock Recirculation of germinal center {B} cells: a multilevel selection
  strategy for antibody maturation.
\newblock Immunol Rev. 2007 1~Apr;216(1):130--141.
\newblock Available from:
  \url{http://dx.doi.org/10.1111/j.1600-065X.2007.00507.x}.

\bibitem{McHeyzer-Williams2015-un}
McHeyzer-Williams LJ, Milpied PJ, Okitsu SL, McHeyzer-Williams MG.
\newblock Class-switched memory {B} cells remodel {BCRs} within secondary
  germinal centers.
\newblock Nat Immunol. 2015 Mar;16(3):296--305.
\newblock Available from: \url{http://dx.doi.org/10.1038/ni.3095}.

\bibitem{Gupta2017-cg}
Gupta NT, Adams KD, Briggs AW, Timberlake SC, Vigneault F, Kleinstein SH.
\newblock Hierarchical Clustering Can Identify {B} Cell Clones with High
  Confidence in Ig Repertoire Sequencing Data.
\newblock J Immunol. 2017 Feb;Available from:
  \url{http://dx.doi.org/10.4049/jimmunol.1601850}.

\bibitem{Wu2011-yy}
Wu X, Zhou T, Zhu J, Zhang B, Georgiev I, Wang C, et~al.
\newblock Focused evolution of {HIV-1} neutralizing antibodies revealed by
  structures and deep sequencing.
\newblock Science. 2011 Aug;333(6049):1593--1602.
\newblock Available from: \url{http://dx.doi.org/10.1126/science.1207532}.

\bibitem{Saitou1987-sr}
Saitou N, Nei M.
\newblock The neighbor-joining method: a new method for reconstructing
  phylogenetic trees.
\newblock Mol Biol Evol. 1987 Jul;4(4):406--425.
\newblock Available from: \url{http://www.ncbi.nlm.nih.gov/pubmed/3447015}.

\bibitem{Gascuel2006-gb}
Gascuel O, Steel M.
\newblock {Neighbor-Joining} Revealed.
\newblock Mol Biol Evol. 2006 Nov;23(11):1997--2000.
\newblock Available from:
  \url{http://mbe.oxfordjournals.org/content/23/11/1997.abstract}.

\bibitem{Doucet2001-av}
Doucet A, de~Freitas N, Gordon N.
\newblock An Introduction to Sequential Monte Carlo Methods.
\newblock In: Sequential Monte Carlo Methods in Practice. Statistics for
  Engineering and Information Science. Springer New York; 2001. p. 3--14.
\newblock Available from: \url{http://dx.doi.org/10.1007/978-1-4757-3437-9_1}.

\bibitem{Laserson2012-pi}
Laserson J.
\newblock Bayesian assembly of reads from high throughput sequencing.
\newblock Stanford; 2012.
\newblock Available from: \url{http://purl.stanford.edu/xp796hy4748}.

\bibitem{Barak2008-fw}
Barak M, Zuckerman N, Edelman H, Unger R, Mehr R.
\newblock {IgTree} (c) : Creating Immunoglobulin variable region gene lineage
  trees.
\newblock Journal of Immunological Methods. 2008;338(1-2):67--74.
\newblock Available from: \url{http://dx.doi.org/10.1016/j.jim.2008.06.006}.

\bibitem{Stern2014-ph}
Stern JNH, Yaari G, Vander~Heiden JA, Church G, Donahue WF, Hintzen RQ, et~al.
\newblock {B} cells populating the multiple sclerosis brain mature in the
  draining cervical lymph nodes.
\newblock Sci Transl Med. 2014 6~Aug;6(248):248ra107.
\newblock Available from: \url{http://dx.doi.org/10.1126/scitranslmed.3008879}.

\bibitem{Felsenstein1978-rr}
Felsenstein J.
\newblock Cases in which Parsimony or Compatibility Methods Will be Positively
  Misleading.
\newblock Syst Zool. 1978 Dec;27(4):401--410.
\newblock Available from: \url{http://www.jstor.org/stable/2412923}.

\bibitem{DeWitt2018-el}
DeWitt WS, Mesin L, Victora GD, Minin VN, Matsen FA IV.
\newblock Using genotype abundance to improve phylogenetic inference.
\newblock Mol Biol Evol. 2018 Feb;Available from:
  \url{https://academic.oup.com/mbe/advance-article/doi/10.1093/molbev/msy020/4893244}.

\bibitem{Kepler2013-sy}
Kepler TB.
\newblock Reconstructing a B-cell clonal lineage. I. Statistical inference of
  unobserved ancestors.
\newblock F1000Res. 2013 Apr;2:103.
\newblock Available from:
  \url{http://dx.doi.org/10.12688/f1000research.2-103.v1}.

\bibitem{Hoehn2017-ol}
Hoehn KB, Lunter G, Pybus OG.
\newblock A Phylogenetic Codon Substitution Model for Antibody Lineages.
\newblock Genetics. 2017 May;206(1):417--427.
\newblock Available from: \url{http://dx.doi.org/10.1534/genetics.116.196303}.

\bibitem{Zhu2013-ax}
Zhu J, Wu X, Zhang B, McKee K, O’Dell S, Soto C, et~al.
\newblock De novo identification of {VRC01} class {HIV-1--neutralizing}
  antibodies by next-generation sequencing of B-cell transcripts.
\newblock Proceedings of the National Academy of Sciences. 2013
  Oct;110(43):E4088--E4097.
\newblock Available from: \url{http://dx.doi.org/10.1073/pnas.1306262110}.

\bibitem{Doria-Rose2014-vi}
Doria-Rose NA, Schramm CA, Gorman J, Moore PL, Bhiman JN, DeKosky BJ, et~al.
\newblock Developmental pathway for potent {V1V2-directed} {HIV-neutralizing}
  antibodies.
\newblock Nature. 2014 Mar;Available from:
  \url{http://dx.doi.org/10.1038/nature13036}.

\bibitem{Felsenstein1989-dt}
Felsenstein J.
\newblock {PHYLIP-Phylogeny} Inference Package (Version 3.2).
\newblock Cladistics. 1989;5:164--166.
\newblock Available from:
  \url{http://www.trex.uqam.ca/index.php?action=phylip&app=promlk}.

\bibitem{Ashkenazy2012-vk}
Ashkenazy H, Penn O, Doron-Faigenboim A, Cohen O, Cannarozzi G, Zomer O, et~al.
\newblock {FastML}: a web server for probabilistic reconstruction of ancestral
  sequences.
\newblock Nucleic Acids Res. 2012 Jul;40(Web Server issue):W580--4.
\newblock Available from: \url{http://dx.doi.org/10.1093/nar/gks498}.

\bibitem{Nguyen2015-bs}
Nguyen LT, Schmidt HA, von Haeseler A, Minh BQ.
\newblock {IQ-TREE}: a fast and effective stochastic algorithm for estimating
  maximum-likelihood phylogenies.
\newblock Mol Biol Evol. 2015 Jan;32(1):268--274.
\newblock Available from: \url{http://dx.doi.org/10.1093/molbev/msu300}.

\bibitem{Hershberg2008-rp}
Hershberg U, Uduman M, Shlomchik MJ, Kleinstein SH.
\newblock Improved methods for detecting selection by mutation analysis of Ig
  {V} region sequences.
\newblock Int Immunol. 2008 Apr;20(5):683--694.
\newblock Available from: \url{http://dx.doi.org/10.1093/intimm/dxn026}.

\bibitem{Uduman2011-ib}
Uduman M, Yaari G, Hershberg U, Stern JA, Shlomchik MJ, Kleinstein SH.
\newblock Detecting selection in immunoglobulin sequences.
\newblock Nucleic Acids Res. 2011 Jun;39(Web Server issue):W499--504.
\newblock Available from: \url{http://dx.doi.org/10.1093/nar/gkr413}.

\bibitem{Yaari2015-ss}
Yaari G, Benichou JIC, Vander~Heiden JA, Kleinstein SH, Louzoun Y.
\newblock The mutation patterns in B-cell immunoglobulin receptors reflect the
  influence of selection acting at multiple time-scales.
\newblock Philos Trans R Soc Lond B Biol Sci. 2015 Sep;370(1676).
\newblock Available from: \url{http://dx.doi.org/10.1098/rstb.2014.0242}.

\bibitem{Dunn-Walters2002-cu}
Dunn-Walters DK, Belelovsky A, Edelman H, Banerjee M, Mehr R.
\newblock The dynamics of germinal centre selection as measured by
  graph-theoretical analysis of mutational lineage trees.
\newblock Dev Immunol. 2002 Dec;9(4):233--243.
\newblock Available from: \url{http://www.ncbi.nlm.nih.gov/pubmed/15144020}.

\bibitem{Uduman2014-pb}
Uduman M, Shlomchik MJ, Vigneault F, Church GM, Kleinstein SH.
\newblock Integrating {B} Cell Lineage Information into Statistical Tests for
  Detecting Selection in Ig Sequences.
\newblock J Immunol. 2014 Feb;192(3):867--874.
\newblock Available from: \url{http://dx.doi.org/10.4049/jimmunol.1301551}.

\bibitem{Liberman2016-qh}
Liberman G, Benichou JIC, Maman Y, Glanville J, Alter I, Louzoun Y.
\newblock Estimate of within population incremental selection through branch
  imbalance in lineage trees.
\newblock Nucleic Acids Res. 2016 Mar;44(5):e46.
\newblock Available from: \url{http://dx.doi.org/10.1093/nar/gkv1198}.

\bibitem{Horns2017-vn}
Horns F, Vollmers C, Dekker CL, Quake SR.
\newblock Signatures of Selection in the Human Antibody Repertoire: Selective
  Sweeps, Competing Subclones, and Neutral Drift; 2017.
\newblock Available from:
  \url{https://www.biorxiv.org/content/early/2017/10/19/145052}.

\bibitem{Neher2014-vl}
Neher RA, Russell CA, Shraiman BI.
\newblock Predicting evolution from the shape of genealogical trees.
\newblock Elife. 2014 Nov;3.
\newblock Available from: \url{http://dx.doi.org/10.7554/eLife.03568}.

\bibitem{Fay2000-qw}
Fay JC, Wu CI.
\newblock Hitchhiking Under Positive Darwinian Selection.
\newblock Genetics. 2000 Jul;155(3):1405--1413.
\newblock Available from: \url{http://www.genetics.org/content/155/3/1405}.

\bibitem{Shlomchik1998-vb}
Shlomchik MJ, Watts P, Weigert MG, Litwin S.
\newblock Clone: A {Monte-Carlo} Computer Simulation of {B} Cell Clonal
  Expansion, Somatic Mutation, and {Antigen-Driven} Selection.
\newblock In: {Professor Garnett Kelsoe}, Flajnik PMF, editors. Somatic
  Diversification of Immune Responses. Current Topics in Microbiology and
  Immunology. Springer Berlin Heidelberg; 1998. p. 173--197.
\newblock Available from:
  \url{http://link.springer.com/chapter/10.1007/978-3-642-71984-4_13}.

\bibitem{Kim2009-dr}
Kim PS, Levy D, Lee PP.
\newblock Modeling and Simulation of the Immune System as a {Self-Regulating}
  Network.
\newblock In: Methods in Enzymology. vol. 467. Academic Press; 2009. p.
  79--109.
\newblock Available from:
  \url{http://www.sciencedirect.com/science/article/pii/S007668790967004X}.

\bibitem{Kleinstein2001-mk}
Kleinstein SH, Singh JP.
\newblock Toward Quantitative Simulation of Germinal Center Dynamics:
  Biological and Modeling Insights from Experimental Validation.
\newblock J Theor Biol. 2001 Aug;211(3):253--275.
\newblock Available from: \url{http://dx.doi.org/10.1006/jtbi.2001.2344}.

\bibitem{Mehr2003-jz}
Mehr R.
\newblock Asynchronous differentiation models explain bone marrow labeling
  kinetics and predict reflux between the pre- and immature {B} cell pools.
\newblock Int Immunol. 2003 1~Mar;15(3):301--312.
\newblock Available from:
  \url{https://academic.oup.com/intimm/article-lookup/doi/10.1093/intimm/dxg025}.

\bibitem{Shahaf2004-kd}
Shahaf G, Allman D, Cancro MP, Mehr R.
\newblock Screening of alternative models for transitional {B} cell maturation.
\newblock Int Immunol. 2004;16(8):1081--1090.
\newblock Available from:
  \url{http://intimm.oxfordjournals.org/content/16/8/1081.full}.

\bibitem{Shahaf2006-fn}
Shahaf G, Johnson K, Mehr R.
\newblock {B} cell development in aging mice: lessons from mathematical
  modeling.
\newblock Int Immunol. 2006 Jan;18(1):31--39.
\newblock Available from: \url{http://dx.doi.org/10.1093/intimm/dxh346}.

\bibitem{Shahaf2010-na}
Shahaf G, Cancro MP, Mehr R.
\newblock Kinetic Modeling Reveals a Common Death Niche for Newly Formed and
  Mature {B} Cells.
\newblock PLoS One. 2010 2~Mar;5(3):e9497.
\newblock Available from:
  \url{http://journals.plos.org/plosone/article/file?id=10.1371/journal.pone.0009497&type=printable}.

\bibitem{Childs2015-cn}
Childs LM, Baskerville EB, Cobey S.
\newblock Trade-offs in antibody repertoires to complex antigens.
\newblock Philos Trans R Soc Lond B Biol Sci. 2015 Sep;370(1676).
\newblock Available from: \url{http://dx.doi.org/10.1098/rstb.2014.0245}.

\bibitem{Wang2015-os}
Wang S, Mata-Fink J, Kriegsman B, Hanson M, Irvine DJ, Eisen HN, et~al.
\newblock Manipulating the selection forces during affinity maturation to
  generate cross-reactive {HIV} antibodies.
\newblock Cell. 2015 Feb;160(4):785--797.
\newblock Available from: \url{http://dx.doi.org/10.1016/j.cell.2015.01.027}.

\bibitem{Wang2017-qu}
Wang S.
\newblock Optimal Sequential Immunization Can Focus Antibody Responses against
  Diversity Loss and Distraction.
\newblock PLoS Comput Biol. 2017 Jan;13(1):e1005336.
\newblock Available from: \url{http://dx.doi.org/10.1371/journal.pcbi.1005336}.

\bibitem{Amitai2017-ku}
Amitai A, Mesin L, Victora G, Kardar M, Chakraborty A.
\newblock A population dynamics model for clonal diversity in a germinal
  center.
\newblock Front Microbiol. 2017;8:1693.
\newblock Available from:
  \url{http://journal.frontiersin.org/article/10.3389/fmicb.2017.01693}.

\bibitem{Liao2013-cr}
Liao HX, Lynch R, Zhou T, Gao F, Alam SM, Boyd SD, et~al.
\newblock Co-evolution of a broadly neutralizing {HIV-1} antibody and founder
  virus.
\newblock Nature. 2013 Apr;496(7446):469--476.
\newblock Available from: \url{http://dx.doi.org/10.1038/nature12053}.

\bibitem{Luo2015-qv}
Luo S, Perelson AS.
\newblock The challenges of modelling antibody repertoire dynamics in {HIV}
  infection.
\newblock Philos Trans R Soc Lond B Biol Sci. 2015 Sep;370(1676).
\newblock Available from: \url{http://dx.doi.org/10.1098/rstb.2014.0247}.

\bibitem{Hershberg2015-hp}
Hershberg U, Luning~Prak ET.
\newblock The analysis of clonal expansions in normal and autoimmune {B} cell
  repertoires.
\newblock Philos Trans R Soc Lond B Biol Sci. 2015 Sep;370(1676).
\newblock Available from: \url{http://dx.doi.org/10.1098/rstb.2014.0239}.

\bibitem{Sok2013-td}
Sok D, Laserson U, Laserson J, Liu Y, Vigneault F, Julien JP, et~al.
\newblock The effects of somatic hypermutation on neutralization and binding in
  the {PGT121} family of broadly neutralizing {HIV} antibodies.
\newblock PLoS Pathog. 2013 Nov;9(11):e1003754.
\newblock Available from: \url{http://dx.doi.org/10.1371/journal.ppat.1003754}.

\bibitem{De_Koning2010-zc}
de~Koning APJ, Gu W, Pollock DD.
\newblock Rapid likelihood analysis on large phylogenies using partial sampling
  of substitution histories.
\newblock Mol Biol Evol. 2010 Feb;27(2):249--265.
\newblock Available from: \url{http://dx.doi.org/10.1093/molbev/msp228}.

\bibitem{Chib1995-ry}
Chib S.
\newblock Marginal Likelihood from the Gibbs Output.
\newblock J Am Stat Assoc. 1995;90(432):1313--1321.
\newblock Available from:
  \url{http://amstat.tandfonline.com/doi/abs/10.1080/01621459.1995.10476635}.

\bibitem{Drummond2005-ks}
Drummond AJ, Rambaut A, Shapiro B, Pybus OG.
\newblock Bayesian coalescent inference of past population dynamics from
  molecular sequences.
\newblock Mol Biol Evol. 2005 Feb;22(5):1185--1192.
\newblock Available from: \url{http://dx.doi.org/10.1093/molbev/msi103}.

\bibitem{Perelson1979-fp}
Perelson AS, Oster GF.
\newblock Theoretical studies of clonal selection: Minimal antibody repertoire
  size and reliability of self-non-self discrimination.
\newblock J Theor Biol. 1979 Dec;81(4):645--670.
\newblock Available from: \url{http://dx.doi.org/10.1016/0022-5193(79)90275-3}.

\bibitem{Hwang2015-wt}
Hwang JK, Alt FW, Yeap LS.
\newblock Related Mechanisms of Antibody Somatic Hypermutation and Class Switch
  Recombination.
\newblock Microbiol Spectr. 2015 Feb;3(1):MDNA3--0037--2014.
\newblock Available from:
  \url{http://dx.doi.org/10.1128/microbiolspec.MDNA3-0037-2014}.

\bibitem{Jackson2014-mo}
Jackson KJL, Wang Y, Collins AM.
\newblock Human immunoglobulin classes and subclasses show variability in {VDJ}
  gene mutation levels.
\newblock Immunol Cell Biol. 2014 Sep;92(8):729--733.
\newblock Available from: \url{http://dx.doi.org/10.1038/icb.2014.44}.

\bibitem{Horns2016-nk}
Horns F, Vollmers C, Croote D, Mackey SF, Swan GE, Dekker CL, et~al.
\newblock Lineage tracing of human {B} cells reveals the in vivo landscape of
  human antibody class switching.
\newblock Elife. 2016 Aug;5.
\newblock Available from: \url{http://dx.doi.org/10.7554/eLife.16578}.

\bibitem{Looney2016-fc}
Looney TJ, Lee JY, Roskin KM, Hoh RA, King J, Glanville J, et~al.
\newblock Human B-cell isotype switching origins of {IgE}.
\newblock J Allergy Clin Immunol. 2016 Feb;137(2):579--586.e7.
\newblock Available from: \url{http://dx.doi.org/10.1016/j.jaci.2015.07.014}.

\bibitem{Pagel1994-ab}
Pagel M.
\newblock Detecting correlated evolution on phylogenies: a general method for
  the comparative analysis of discrete characters.
\newblock Proc R Soc Lond B Biol Sci. 1994 Jan;255(1342):37--45.
\newblock Available from:
  \url{http://rspb.royalsocietypublishing.org/content/255/1342/37}.

\bibitem{Pagel2004-ex}
Pagel M, Meade A, Barker D.
\newblock Bayesian estimation of ancestral character states on phylogenies.
\newblock Syst Biol. 2004 Oct;53(5):673--684.
\newblock Available from: \url{http://dx.doi.org/10.1080/10635150490522232}.

\bibitem{Kaplinsky2016-gt}
Kaplinsky J, Arnaout R.
\newblock Robust estimates of overall immune-repertoire diversity from
  high-throughput measurements on samples.
\newblock Nat Commun. 2016 Jun;7:11881.
\newblock Available from: \url{http://dx.doi.org/10.1038/ncomms11881}.

\bibitem{Meng2017-im}
Meng W, Zhang B, Schwartz GW, Rosenfeld AM, Ren D, Thome JJC, et~al.
\newblock An atlas of B-cell clonal distribution in the human body.
\newblock Nat Biotechnol. 2017 Aug;Available from:
  \url{http://dx.doi.org/10.1038/nbt.3942}.

\bibitem{Havenar-Daughton2016-vk}
Havenar-Daughton C, Carnathan DG, Torrents de~la Pe{\~n}a A, Pauthner M, Briney
  B, Reiss SM, et~al.
\newblock Direct Probing of Germinal Center Responses Reveals Immunological
  Features and Bottlenecks for Neutralizing Antibody Responses to {HIV} Env
  Trimer.
\newblock Cell Rep. 2016 Nov;17(9):2195--2209.
\newblock Available from: \url{http://dx.doi.org/10.1016/j.celrep.2016.10.085}.

\bibitem{Wine2013-ux}
Wine Y, Boutz DR, Lavinder JJ, Miklos AE, Hughes RA, Hoi KH, et~al.
\newblock Molecular deconvolution of the monoclonal antibodies that comprise
  the polyclonal serum response.
\newblock Proc Natl Acad Sci U S A. 2013 Feb;110(8):2993--2998.
\newblock Available from: \url{http://dx.doi.org/10.1073/pnas.1213737110}.

\bibitem{Lavinder2014-tx}
Lavinder JJ, Wine Y, Giesecke C, Ippolito GC, Horton AP, Lungu OI, et~al.
\newblock Identification and characterization of the constituent human serum
  antibodies elicited by vaccination.
\newblock Proc Natl Acad Sci U S A. 2014 Jan;Available from:
  \url{http://dx.doi.org/10.1073/pnas.1317793111}.

\bibitem{Nivala2013-ic}
Nivala J, Marks DB, Akeson M.
\newblock Unfoldase-mediated protein translocation through an
  $\alpha$-hemolysin nanopore.
\newblock Nat Biotechnol. 2013 Feb;31(3):nbt.2503.
\newblock Available from: \url{http://www.nature.com/articles/nbt.2503}.

\bibitem{Doolan2008-xw}
Doolan DL, Mu Y, Unal B, Sundaresh S, Hirst S, Valdez C, et~al.
\newblock Profiling humoral immune responses to P. falciparum infection with
  protein microarrays.
\newblock Proteomics. 2008 Nov;8(22):4680--4694.
\newblock Available from: \url{http://dx.doi.org/10.1002/pmic.200800194}.

\bibitem{Hertz2017-gt}
Hertz T, Beatty PR, MacMillen Z, Killingbeck SS, Wang C, Harris E.
\newblock Antibody Epitopes Identified in Critical Regions of Dengue Virus
  Nonstructural 1 Protein in Mouse Vaccination and Natural Human Infections.
\newblock J Immunol. 2017 May;198(10):4025--4035.
\newblock Available from: \url{http://dx.doi.org/10.4049/jimmunol.1700029}.

\bibitem{Gee2017-dt}
Gee MH, Han A, Lofgren SM, Beausang JF, Mendoza JL, Birnbaum ME, et~al.
\newblock Antigen Identification for Orphan {T} Cell Receptors Expressed on
  {Tumor-Infiltrating} Lymphocytes.
\newblock Cell. 2017 Dec;0(0).
\newblock Available from:
  \url{http://www.cell.com/article/S0092867417314332/abstract}.

\bibitem{Smith2004-ub}
Smith DJ, Lapedes AS, de~Jong JC, Bestebroer TM, Rimmelzwaan GF, Osterhaus
  ADME, et~al.
\newblock Mapping the antigenic and genetic evolution of influenza virus.
\newblock Science. 2004 Jul;305(5682):371--376.
\newblock Available from: \url{http://dx.doi.org/10.1126/science.1097211}.

\bibitem{Katzelnick2015-rr}
Katzelnick LC, Fonville JM, Gromowski GD, Bustos~Arriaga J, Green A, James SL,
  et~al.
\newblock Dengue viruses cluster antigenically but not as discrete serotypes.
\newblock Science. 2015 Sep;349(6254):1338--1343.
\newblock Available from: \url{http://dx.doi.org/10.1126/science.aac5017}.

\bibitem{Francis1960-nz}
Francis T Jr.
\newblock On the Doctrine of Original Antigenic Sin.
\newblock Proc Am Philos Soc. 1960 Dec;104(6):572--578.
\newblock Available from: \url{http://www.jstor.org/stable/985534}.

\bibitem{Andrews2015-ls}
Andrews SF, Huang Y, Kaur K, Popova LI, Ho IY, Pauli NT, et~al.
\newblock Immune history profoundly affects broadly protective {B} cell
  responses to influenza.
\newblock Sci Transl Med. 2015 Dec;7(316):316ra192.
\newblock Available from: \url{http://dx.doi.org/10.1126/scitranslmed.aad0522}.

\bibitem{Horwitz2017-cz}
Horwitz JA, Bar-On Y, Lu CL, Fera D, Lockhart AAK, Lorenzi JCC, et~al.
\newblock Non-neutralizing Antibodies Alter the Course of {HIV-1} Infection In
  Vivo.
\newblock Cell. 2017 Aug;170(4):637--648.e10.
\newblock Available from: \url{http://dx.doi.org/10.1016/j.cell.2017.06.048}.

\bibitem{Luksza2014-xf}
{\L}uksza M, L{\"a}ssig M.
\newblock A predictive fitness model for influenza.
\newblock Nature. 2014 Feb;Available from:
  \url{http://dx.doi.org/10.1038/nature13087}.

\bibitem{Neher2016-mn}
Neher RA, Bedford T, Daniels RS, Russell CA, Shraiman BI.
\newblock Prediction, dynamics, and visualization of antigenic phenotypes of
  seasonal influenza viruses.
\newblock Proc Natl Acad Sci U S A. 2016 Mar;113(12):E1701--9.
\newblock Available from: \url{http://dx.doi.org/10.1073/pnas.1525578113}.

\bibitem{Zarnitsyna2015-cn}
Zarnitsyna VI, Ellebedy AH, Davis C, Jacob J, Ahmed R, Antia R.
\newblock Masking of antigenic epitopes by antibodies shapes the humoral immune
  response to influenza.
\newblock Philos Trans R Soc Lond B Biol Sci. 2015 Sep;370(1676).
\newblock Available from: \url{http://dx.doi.org/10.1098/rstb.2014.0248}.

\bibitem{Zarnitsyna2016-ev}
Zarnitsyna VI, Lavine J, Ellebedy A, Ahmed R, Antia R.
\newblock Multi-epitope Models Explain How Pre-existing Antibodies Affect the
  Generation of Broadly Protective Responses to Influenza.
\newblock PLoS Pathog. 2016 Jun;12(6):e1005692.
\newblock Available from: \url{http://dx.doi.org/10.1371/journal.ppat.1005692}.

\bibitem{Doud2017-fj}
Doud MB, Hensley SE, Bloom JD.
\newblock Complete mapping of viral escape from neutralizing antibodies.
\newblock PLoS Pathog. 2017 Mar;13(3):e1006271.
\newblock Available from: \url{http://dx.doi.org/10.1371/journal.ppat.1006271}.

\bibitem{Dingens2017-ar}
Dingens AS, Haddox HK, Overbaugh J, Bloom JD.
\newblock Comprehensive Mapping of {HIV-1} Escape from a Broadly Neutralizing
  Antibody.
\newblock Cell Host Microbe. 2017 Jun;21(6):777--787.e4.
\newblock Available from: \url{http://dx.doi.org/10.1016/j.chom.2017.05.003}.

\bibitem{Doud2018-oi}
Doud MB, Lee JM, Bloom JD.
\newblock Quantifying the effects of single mutations on viral escape from
  broad and narrow antibodies to an {H1} influenza hemagglutinin.
\newblock bioRxiv. 2018;Available from:
  \url{https://www.biorxiv.org/content/early/2018/01/22/210468.abstract}.

\bibitem{Forsyth2013-zh}
Forsyth CM, Juan V, Akamatsu Y, DuBridge RB, Doan M, Ivanov AV, et~al.
\newblock Deep mutational scanning of an antibody against epidermal growth
  factor receptor using mammalian cell display and massively parallel
  pyrosequencing.
\newblock MAbs. 2013 Jul;5(4):523--532.
\newblock Available from: \url{http://dx.doi.org/10.4161/mabs.24979}.

\bibitem{Adams2016-ja}
Adams RM, Mora T, Walczak AM, Kinney JB.
\newblock Measuring the sequence-affinity landscape of antibodies with
  massively parallel titration curves.
\newblock Elife. 2016 Dec;5.
\newblock Available from: \url{http://dx.doi.org/10.7554/eLife.23156}.

\bibitem{Fonville2014-rl}
Fonville JM, Wilks SH, James SL, Fox A, Ventresca M, Aban M, et~al.
\newblock Antibody landscapes after influenza virus infection or vaccination.
\newblock Science. 2014 Nov;346(6212):996--1000.
\newblock Available from: \url{https://dx.doi.org/10.1126/science.1256427}.

\bibitem{Brodin2017-qq}
Brodin P, Davis MM.
\newblock Human immune system variation.
\newblock Nat Rev Immunol. 2017 Jan;17(1):21--29.
\newblock Available from: \url{http://dx.doi.org/10.1038/nri.2016.125}.

\end{thebibliography}

\end{document}